\documentclass[final]{IEEEtran}
\IEEEoverridecommandlockouts
\usepackage{color}

\usepackage{times}
\usepackage[final]{graphicx}
\usepackage{amsmath,amsfonts}
\usepackage{amsthm}
\usepackage{amssymb,amsbsy}

\usepackage{cite}
\usepackage{multirow}

\usepackage{algorithm}
\usepackage{algpseudocode}

\newcommand{\ignore}[1]{}

\newcommand{\indic}{\mbox{$1\!\!1$}}

\newcommand{\hspp}{\hspace{0.05in} }
\newcommand{\hsppp}{\hspace{0.02in} }

\newsavebox{\savepar}

\begin{document}
\title{Antenna Placement and Performance Tradeoffs \\
with Hand Blockage in Millimeter Wave Systems}
\author{\normalsize Vasanthan Raghavan, Mei-Li (Clara) Chi, M.\ Ali Tassoudji,
Ozge H.\ Koymen, and Junyi Li
\thanks{The authors are with Qualcomm Flarion Technologies, Bridgewater, NJ 08807, USA
and Qualcomm Corporate R\&D, San Diego, CA 92121, USA.
Email: \tt{vasanthan\_raghavan@ieee.org}, {\tt clarachi@gmail.com}, {\tt \{alit,okoymen,junyil\}@qti.qualcomm.com.}}
}

\maketitle
\vspace{-10mm}

\begin{abstract}
\noindent
The ongoing commercial deployment of millimeter wave systems brings into focus a
number of practical issues in form factor user equipment (UE) design. With
wavelengths becoming smaller, antenna gain patterns becoming directional, and
link budgets critically dependent on beamforming, it becomes imperative to
use a number of antenna modules at different locations of the UE for good
performance. While more antennas/modules can enhance beamforming array gains, it
comes with the tradeoff of higher component cost, power consumption of the associated
radio frequency circuitry, and a beam management overhead in learning the appropriate beam
weights. Thus, the goal of a good UE design is to provide robust {\em spherical coverage}
corresponding to good array gains over the entire sphere around the UE with a low beam
management overhead, complexity, and cost. The scope of this paper is to study the
implications of two popular commercial millimeter wave UE designs (a {\em face} and an
{\em edge} design) on spherical coverage. We show that analog beam codebooks can result
in good performance for both the designs, and the edge design provides a better tradeoff
in terms of robust performance (with hand blockage), beam management overhead, implementation
complexity from an antenna placement standpoint and cost.
\end{abstract}

\begin{keywords}
\noindent Millimeter wave, commercial deployments, spherical coverage,
antenna placement, modular design, UE design, 5G-New Radio,
hand blockage. 
\end{keywords}

\section{Introduction}
\label{sec1}
Enhanced spectrum availability over a part of the millimeter wave band
($\sim{\hspace{-0.03in}}24$-$100$ GHz) has led to the focus of Fifth Generation
(5G) wireless systems at these bands to meet the increased data rate and low
latency requirements~\cite{khan,qualcomm,rappaport,boccardi1}. With the ongoing
standardization and testing of such systems, a number of practical issues have to
be solved before (and through) commercial deployments. In this context, it is
now well-understood that millimeter wave link margins are sufficient to allow
small-to-medium cell coverage by leveraging the increased beamforming array gains
possible with the use of a larger number of antennas within the same physical
aperture~\cite{3gpp_CM_rel14_38901,5G_whitepaper,802d11_maltsev,802d11ay_maltsev,metis2020,vasanth_tap2018,rappaport_chmodel,shafi_5G}.
In contrast to sub-$6$ GHz systems, such array gains are constrained by the fact
that an antenna at millimeter wave carrier frequencies is inherently
directional\footnote{In particular, a millimeter wave antenna can produce meaningful
gains only over a certain spatial coverage region (typically a $120^{\sf o} \times
120^{\sf o}$ region of the sphere).}. As with legacy systems, the base-station design
can naturally incorporate this constraint by focussing on realizing sectoral coverage
(typically a $120^{\sf o}$ or $90^{\sf o}$ region in the azimuth plane, with a narrow
$30^{\sf o}$-$45^{\sf o}$ elevation steering/coverage). However, a similar design
objective at the user equipment (UE) end can lead to significant performance degradation
if useful signals cannot be
picked up from different base-stations serving in different sectors (individually or
in coordination) or from different clusters (within the same base-station) that
correspond to widely disparate angles of arrival. Thus, a cumulative distribution
function (CDF) of the {\em spherical coverage} that captures the beamforming array
gain achievable with the UE's antennas in a sphere ($360^{\sf o} \times 180^{\sf o}$
in azimuth and elevation, respectively) around it becomes a paramount benchmark of UE
design and performance.

In particular, a good spherical coverage CDF corresponds to {\em not only} good
array gains in the top few (e.g., top $30$) percentile points, but also in the middle
(e.g., $30^{\sf th}$-$75^{\sf th}$) percentile points. In particular, as we will see
later in the sequel, the hand can lead to a wide spatial area blockage and good
coverage cannot be ensured with any UE design over the tail (approximately bottom
$25$) percentile points. With only directional/modular coverage possible with
millimeter wave antennas, it becomes necessary to use multi-antenna subarrays at
different locations of the UE to realize a good spherical coverage CDF. Such a
construction becomes even more critical to provide subarray diversity and robustness
to hand blockage.

On the other hand, with a reduction in wavelength, a number of individual
antenna elements can be placed/mounted at the UE side within the same form factor
allowing increased array gains that is hitherto not possible at sub-$6$ GHz. While
such a possibility makes a theoretical case for packing as many antennas as possible
at the UE side (contingent on competing space with antennas at sub-$6$ GHz
frequencies, WiFi and Bluetooth systems, cameras, sensors, and the associated circuit
elements, etc.), the added cost of
millimeter wave antenna modules and associated radio frequency (RF) front end circuitry (e.g., power
and low-noise amplifiers, mixers, etc.) and the concomitant power increase puts a
practical limit on how many antennas can be gainfully employed in a millimeter wave
UE. More importantly, while the use of a large number of antennas (and antenna modules)
can {\em theoretically} lead to increased beamforming gains, if these capabilities are
not {\em practically} exercisable with a low beam management overhead, the capabilities
can quickly turn out to be onerous and a curse rather than a blessing.

With form factor/real-estate considerations at the UE side, the focus of this work
is on the practical realm: antenna placement and the impact of different UE designs on
spherical coverage,
both with and without hand blockage. To understand the tradeoffs in terms of antenna
placement, we consider two popular UE designs in this work. These are: i) a {\em face}
design that has antenna modules, with planar dual-polarized patch subarrays and linear
dipole subarrays on the edges, placed on the front and back faces of the UE and
ii) an {\em edge} design with linear dual-polarized patch subarrays placed on three edges
of the UE. These designs have been introduced/studied as possible commercial UE designs
by different original equipment manufacturers (OEMs) for compliance and testing studies
of millimeter wave systems at the Third Generation Partnership Project (3GPP) Working
Group 4 (WG4) meetings; see
e.g.,~\cite{38_803spec,apple_intel_UEdesign,ss_apple_intel_UEdesign,dcm_UEdesign,qcom_UEdesign,qcom_UEdesign2,sony_UEdesign,sony_UEdesign2,sony_UEdesign3,sony_UEdesign4,sony_ericsson_UEdesign,sony_ericsson_UEdesign2,LGE_UEdesign}, etc.

Due to the complicated impact of different substrate materials in the UE on the antenna
response function, such studies
cannot be conducted theoretically. Thus, for both the UE designs, the individual antenna
element response functions in both polarizations are obtained over the sphere using the
Ansys High-Frequency antenna response/Structure Simulator (HFSS) commercial software
suite~\cite{hfss}. With these individual response functions, array gains corresponding
to different beamforming schemes such as maximum ratio combining (MRC), equal gain
combining (EGC), an RF/analog beam codebook-based solution, and antenna selection are
studied. The MRC and EGC schemes can be viewed as optimistic upper bounds (with different
beam weight constraints) on beamforming performance given a certain antenna/module
capability. The antenna selection scheme can be viewed as a pessimistic lower bound
corresponding to legacy system design. It can also be viewed as the performance obtained
after the initial acquisition phase in 3GPP 5G-New Radio (NR) using\footnote{In 3GPP
TR 38.912~\cite{3gpp_CM_rel14_38912}, hierarchical beamforming at the base-station
and UE ends are proposed in three phases: an initial acquisition or P-1 phase of
wide beams at both ends, and beam refinement (or P-2 and P-3 phases) at the
base-station and UE ends, respectively. The Release 15 spec of 5G-NR essentially
follows these outlines without any explicit citations to P-1/2/3 procedures.} the
P-1 procedure~\cite{3gpp_CM_rel14_38912}
with a single antenna exciting {\em pseudo-omni} beam. On the other hand, the
RF/analog beam codebook scheme is a 3GPP 5G-NR-compatible scheme for realizing
practical beamforming performance in millimeter wave systems at the end of the P-1/2/3
procedures. The size of the beamforming codebook determines the tradeoff between beam
management overhead and gap to optimal MRC/EGC performance. The larger the codebook
size, the higher the beam management overhead and smaller the gap to optimal performance
(and {\em vice versa}).

We consider both Portrait and Landscape mode blockages with blockage model from
the 3GPP specification in TR 38.901~\cite{3gpp_CM_rel14_38901} as well as more
realistic models from measurements with a $28$ GHz experimental prototype~\cite{vasanth_comm_mag_16,vasanth_blockage_tap2018}. Based on these
studies, the main conclusions of this work are as follows.
\begin{itemize}
\item
Good RF/analog beam codebooks that tradeoff beam management overhead for robust
performance over the sphere can be designed for both the UE designs considered
in this work. These codebooks can realize a significant fraction (within $1$-$2$ dB)
of the optimal array gains possible for these designs. Substantial performance
improvement (from $2$-$6$ dB depending on the size of the subarrays and the precise
direction of the cluster/path) is seen with the codebook scheme over the antenna
selection scheme. This performance improvement captures the benefit of performing the
P-3 beam refinement at the UE side over the P-1 initial acquisition phase.

\item
The flat $30$ dB loss over the blocked spatial region assumed with the 3GPP
blockage model in~\cite{3gpp_CM_rel14_38901} leads to an abrupt and dramatic
loss in performance over these regions in both Portrait and Landscape modes. A
more smoother performance degradation is seen with the model proposed
in~\cite{vasanth_blockage_tap2018}. Nevertheless, blockage is seen to produce
a {\em bimodal} behavior of nearly unobstructed transmission/reception over the
unblocked region, and unrecoverable signal over the blocked region. This bimodal
performance reinforces the criticality of subarray/modular diversity, channel
richness, and alternate viable cluster/path learning for switching to a
potentially alternate cluster/path in the case of an impending blockage of a
serving cluster/path~\cite{vasanth_blockage_tap2018,vasanth_comm_mag_18}.

\item
While the use of an increased number of antenna modules would suggest a better
robustness to blockage, the learning overhead associated with codebook-based beam
training suggests a good tradeoff point in terms of the number of antenna modules
at an intermediate value. The face and edge designs are generally competitive with
each other with no strong advantages in performance for either design. In general,
practical advantages can be seen with a smaller number of subarrays/modules to be
learned/trained.

\item
That said, the face design has a strong implementation-level complexity arising from the
need to find real-estate on the front face of the UE (something that is typically
reserved for almost bezel-less displays in current and next generations of UEs).
Further, placing the antenna module underneath a glass/plastic display can lead
to additional signal
deterioration~\cite{vasanth_tap2018,sony_ran4_glassplastic,qcom_ran4_glassplastic}
that is unaccounted for. A third complexity associated with the face design are
transmissions that could cause a major exposure in the direction of unintended body
parts (e.g., eye, skin, etc.). Thus, the edge design provides a better tradeoff in
terms of robust performance, beam management overhead, implementation complexity, and
cost, suggesting its utility in commercial millimeter wave UE designs.
\end{itemize}

\noindent
{\bf \underline{Novelty of this work:}} The novelty of this paper relative to prior
works on beamforming is now explained. While there are a number of works on beamforming
for millimeter wave systems (single- and multi-cell aspects), a system level study of
the tradeoffs in antenna placement in a form factor constrained UE have not been
explored prior to this work. Quite simply, unlike sub-$6$ GHz systems where antenna
placement does not matter much in terms of system level performance, the success of
millimeter wave systems is {\em critically} dependent on good antenna placement and there
are no {\em fair} studies of different UE designs in the literature. Specifically,
notions such as {\em spherical coverage} have not been studied in an academic context
to compare multiple UE designs, either with or without hand blockage. Such studies are
important given the impending commercialization of millimeter wave
systems~\cite{qualcomm_mmw_modules} and the need for the robustness of such designs
with hand blockage. This work focusses on answers to these practically-inspired problems.

In terms of the prior work of the authors,~\cite{vasanth_tap2018} compares
the macroscopic 
channel features (such as path loss exponents, delay spread, penetration loss, etc.)
via measurements in
different deployment scenarios across different carrier frequencies, but limits itself to
channel modeling and its implications. The work in~\cite{vasanth_comm_mag_16} describes a
$28$ GHz experimental prototype with a proprietary pre-5G subframe structure and the
prototype's robustness in terms of performance with
indoor and outdoor mobility. The work in~\cite{vasanth_blockage_tap2018} describes the
limitations of the 3GPP hand blockage model relative to form factor UE measurements with
the hand. To overcome these limitations, it proposes a simplified alternate model describing
the hand blockage loss. Both these models are used in this work to capture spherical coverage
loss with hand blockage. The scope of~\cite{vasanth_comm_mag_18} is a broad summary (for a wider
reach) of the spatio-temporal impact of hand blockage in millimeter wave systems. A brief
explanation of the poor fit of the 3GPP model as illustrated in~\cite{vasanth_blockage_tap2018}
along with the time-scales of blockage and link disruption, UE side impact and possible mitigation
strategies are considered in~\cite{vasanth_comm_mag_18}. In~\cite{raghavan_jstsp}, different types of
beamforming schemes (such as those based on singular vectors, array steering vectors,
compressive sensing schemes, etc.) are studied and a simple codebook-based beamforming
scheme is shown to be robust, practical and scalable for initial link acquisition in
millimeter wave systems. Such studies have also motivated the choice behind the agreed
protocol in 3GPP TR 38.912~\cite{3gpp_CM_rel14_38912}. 

\noindent
{\bf \underline{Organization:}} This paper is organized as follows. Sec.~\ref{sec2}
introduces the UE designs
studied in this work, their design tradeoffs in terms of practical implementation,
and connections to prior UE design work in academia as well as commercial designs.
Sec.~\ref{sec3}
explains the system setup 
such as the nature and scope of the beamforming algorithms considered, performance
metric used in the study for quantifying the goodness of the designs, RF/analog beam
codebooks used for these designs, and
the considered blockage models. Sec.~\ref{sec4} presents the spherical coverage
CDF tradeoffs for these designs (as well as more sophisticated designs) with the
different beamforming schemes and blockage
models as well as head-to-head comparisons across the designs. Sec.~\ref{sec5}
presents some concluding remarks and possible directions for future studies.


\section{UE Designs for Spherical Coverage Studies}
\label{sec2}

\begin{figure}[htb!]
\begin{center}
\includegraphics[height=2.8in,width=3.6in] {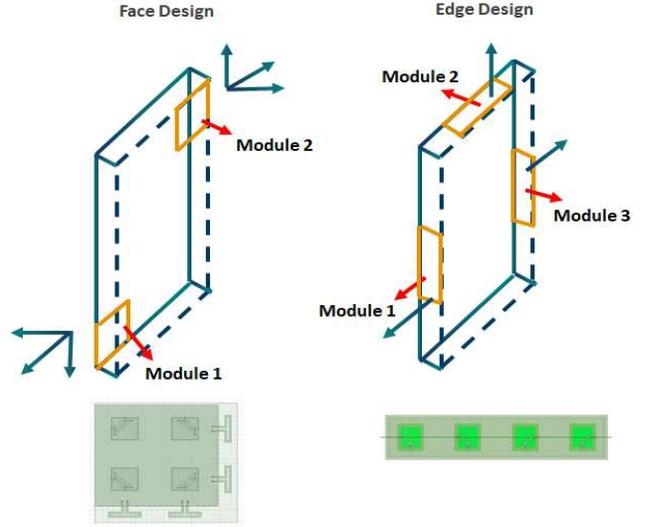}
\caption{\label{fig_UE_Designs}
Pictorial illustration of the {\em face} and {\em edge}
designs considered in this work along with the antenna module structure in these designs
and the boresight directions of the main scanning plane(s) of all the subarrays.}
\end{center}
\end{figure}

\subsection{Designs Studied in This Work}
\label{subsec2_a}
For spherical coverage studies, we consider two popular UE designs in this work as
illustrated in Fig.~\ref{fig_UE_Designs}. These designs are:
\begin{itemize}
\item 
A {\em face} design with two antenna modules (on the front and back
corners of the UE) with each module made of $2 \times 2$ dual-polarized patch subarrays
and two $2 \times 1$ (single-polarized) dipole subarrays at the edges of the module. Note
that it is typical to count a dual-polarized patch antenna as two antenna elements since they
are fed by two independent antenna feeds. Thus, for each antenna module in the face design,
the $2 \times 2$ dual-polarized patch
subarray counts for $8$ antenna elements and the dipole subarrays count for $4$ antenna
elements leading to a total of $12$ antenna elements per module. Since there are two modules,
we have $24$ antenna elements in all in this design.



\item 
An {\em edge} design with three antenna modules (on
three sides of the UE) with each module made of $4 \times 1$ dual-polarized
patch subarrays alone. For each antenna module in the edge design,
the dual-polarized patch subarray accounts for $8$ antenna elements. Along with the use of three
modules, we have $24$ elements in all here.

\end{itemize}
Both these designs are tailored for dual layer transmission/reception at $28$ GHz.
These two layers could be mapped/connected to the two polarizations of the patch
subarrays, or two paired dipole subarrays, or a patch and a dipole subarray (the
latter two pairs arise only in the face design). Information on the number of
antenna modules, number of antennas (in both polarizations), approximate elemental
gains of the antennas, and number and description of the different subarrays 
in these designs are summarized in Table~\ref{table_UE_Design}. A pictorial
illustration of the antenna module design and the boresight directions of the main
scanning plane of all the subarrays in each design are marked with blue arrows in
Fig.~\ref{fig_UE_Designs}.

\begin{table*}[htb!]
\caption{Design Parameters for the Face and Edge Designs}
\label{table_UE_Design}
\begin{center}
\begin{tabular}{|c||c|c|}
\hline
Parameter of interest & Face design & Edge design  \\
\hline
\hline
No.\ of antenna modules & $2$ & $3$ \\ \hline
No.\ of antenna elements & $24$ & $24$ \\ \hline
No.\ of subarrays per module & $4$ ($2 \times 2$ dual-polarized patch subarrays,
& $2$ ($4 \times 1$ dual-polarized patch subarrays) \\
& $2 \times 1$ and $1 \times 2$ dipole subarrays) & \\
\hline
No.\ of beams per subarray & $4$ for patch and $2$ for dipole subarrays
& $4$ for each patch subarray \\ \hline
Codebook size per module & $12$ (= $4$ beams $\times$ $2$ patch subarrays &
$8$ (= $4$ beams $\times$ $2$ patch subarrays) \\
& + $2$ beams $\times$ $2$ dipole subarrays)
&
\\ \hline
Total codebook size & $24 = 12 \times 2$ & $24 = 8 \times 3$ \\ \hline
Elemental gain & $\sim 5.8$ and $\sim 4.7$ dBi for patches and dipoles
& $\sim 5.5$ dBi \\
\hline
\end{tabular}
\end{center}
\end{table*}

\subsection{Design Tradeoffs and Practical Implementation Issues} 
\label{subsec2_b}
\noindent{\bf \underline{Face Design:}} The face design is hard to implement in practice. This
is because in addition to finding sufficient real-estate within the display unit\footnote{Almost
bezel-less displays have become popular in the current generation of UEs and will be increasingly
used in future designs. This constraint renders the use of a planar array at least on the
front face of the UE questionable.} of the UE for the antenna module to be mounted,
careful mounting of the antenna modules can also lead to manufacturing complexities
and cost/time overruns. Further, such a design can incur significant additional
radiation losses due to penetration of millimeter wave signals through typical display
materials (e.g., glass, plastic, ceramic, etc.). In particular, works such
as~\cite{vasanth_tap2018},~\cite{sony_ran4_glassplastic} and~\cite{qcom_ran4_glassplastic}
point out that the loss is material (permittivity and loss tangent)-dependent, depends
on the antenna type (dipole or patch), clearance between display and cover, and can cover
a wide bandwidth. The cover/display acts as a lens/dispersive medium and scatters the signal
compared to the baseline case of no cover. A
glass cover will scatter more energy than a plastic cover resulting in attenuation of
signals in certain directions and comparable performance or even amplification of
signals in certain other directions (all relative to the case with no cover).

Another issue with the use of planar arrays on the front face of the UE is exposure of
sensitive body parts (e.g., eye, skin, etc.) to the beamformed signal with high energy.
On the other hand, the use of planar arrays with each antenna module (instead of linear
arrays) allows a two-dimensional beam scanning that allows a better parsing of the
clusters in the channel, as well as limiting signal leakage (or interference) in
unintended directions possible with one-dimensional beam scanning. Thus, a reasonable
spherical coverage can be anticipated with the use of only two antenna modules (on the
front and back) which can minimize cost, power consumption, as well as beam management
overhead.

Specific to the face design, dipole antennas are more affected by placement issues than
patch antennas. Thus, the beamforming performance with the dipole antennas in these
designs can show a big deviation from expected performance
in Freespace. This deviation requires a careful design of housing in these designs. Further,
dipole antennas require more area (and a bigger size) than patch antennas. Thus, in thin
UE designs, the antenna modules may need to be tilted or placed at an angle resulting in a
more complicated spherical coverage tradeoff. On the other hand, dipoles allow a more broadband
coverage relative to patch antenna elements allowing the reuse of the same antenna
design across different bands/geographies~\cite{sony_ericsson_dipoles}.

\noindent{\bf \underline{Edge Design:}} The edge design appears to be the easiest from a
practical implementation standpoint. By placing antennas on the edges, the edge design
takes advantage of the robustness to the precise choice of location of the antenna modules
on the edge and thus this design can minimize mounting problems. Since a commercial UE
design has to accommodate different real-estate constraints associated with sensors, cameras,
battery, other antennas, etc., this robustness adds a significant level of versatility to
UE design. Additionally, the edge placement can significantly reduce display-related
penetration losses (relative to the face design).

On the other hand, analogous to display-related losses for the face design,
frame-related losses can accrue for the edge design. Note that typical frame materials
include plastic and metal. The typical impact of these materials is to decrease the beam's
strength and/or to tilt or steer the beams away from their intended directions. From prior
works such as~\cite{sony_ran4_glassplastic} and~\cite{qcom_ran4_glassplastic}, it is known
that additional losses are a function of the permittivity and loss tangent of the material,
antenna type, clearance between frame and antenna substrate, beam steering direction, etc.
In the case of plastic frames, commonly used in a broad range of UEs, these losses are
usually minimal. However, metallic frames can lead to further losses that need to be
included in our studies. To be fair to both the face and edge designs, we have not
included display losses for the face design as well as frame losses for the edge design
in our studies. This is done so as to give a big-picture idea of the tradeoffs involved in
UE design instead of incorporating every implementation aspect in high specificity. For
practical implementations, both these additional losses need to be included.

The use of linear arrays (a UE's typical form-factor only allows linear arrays on the edges)
leads to the need of at least four antenna modules for full spherical coverage in Freespace
resulting in increased cost, power consumption as well as beam management overhead. In general,
the edge design tries to appropriate the good features of the face design such as a small number
of antenna modules by adding a layer of robustness in design. By anticipating poor spherical
coverage performance over one part of the sphere, the number of antenna modules can be reduced
from four to three. This poor performance could be due to the edge pointing away from the serving
base-station(s) and towards the ground plane in Portrait mode, or due to the presence of the hand
in the Landscape mode. In terms of exposure, some subarrays can steer energy towards the body of
the user with minor signal energy peaks. Thus, relative to the face design, the edge design is
expected to have rather minor exposure-related concerns. With this background, it is of interest
in understanding the spherical coverage CDF performance with these UE designs.

\subsection{Connections to Other Designs}
\label{subsec2_c}
The readers are pointed to~\cite{ksahota_5Gsummit,tap_overview1,tap_overview2} for some
recent studies on design tradeoffs of 5G antenna arrays with form factor considerations.
In the context of spherical coverage studies, a number of reference UE architecture designs
have been introduced at 3GPP in terms of developing testing and conformance requirement
specifications for the effective isotropically radiated power (EIRP) with millimeter wave
transmissions. For example, the TR 38.803 spec document has a number of potential UE
reference architectures for the high bands
($>24$ GHz)~\cite[Sec.\ 6.2.1.1, pp.\ 107-108]{38_803spec}; also,
see~\cite{ericsson_sony_refUEdesigns}. In the spherical coverage
studies considered at the WG4 level, a number of companies have proposed and considered
diverse UE designs. These designs include the proposals by Apple and
Intel~\cite{apple_intel_UEdesign}, Samsung, Apple and
Intel~\cite{ss_apple_intel_UEdesign}, NTT DOCOMO~\cite{dcm_UEdesign},
Qualcomm~\cite{qcom_UEdesign,qcom_UEdesign2}, Sony~\cite{sony_UEdesign,sony_UEdesign2,sony_UEdesign3,sony_UEdesign4},
Sony and Ericsson~\cite{sony_ericsson_UEdesign,sony_ericsson_UEdesign2},
LG Electronics~\cite{LGE_UEdesign}, etc.

In particular, a design similar
to the face design has been considered in~\cite{38_803spec,ksahota_5Gsummit} as well as
by~\cite{dcm_UEdesign,sony_UEdesign2,sony_UEdesign3,ss_apple_intel_UEdesign,LGE_UEdesign}.
The edge design has been proposed and studied in~\cite{qcom_UEdesign,sony_UEdesign2}. More details
on its practical implementation considerations (such as packaging and performance with glass/plastic
display materials) are discussed in~\cite{qcom_UEdesign2}. 
Many other designs are considered in~\cite{sony_ericsson_UEdesign2,ericsson_sony_refUEdesigns,helander}. Thus,
these designs can be seen to be representative and reflective of popular commercial UE deployments
(current as well as the near-future) in the millimeter wave regime.


Some recent work documenting the robustness advantages of large circular arrays for
outdoor deployments include~\cite{circ_array1}. In this context, in terms of the
directivity pattern, a linear array is more asymmetric due to the one-dimensional
nature of the array relative to a planar array which is two-dimensional. Similarly,
a circular array is more symmetric due to the effective two-dimensional nature of
the array. On the other hand, due to form factor constraints, circular arrays can
only be deployed on the front or back face of the UE instead of the edges. Thus,
circular arrays share the same pros and cons as planar arrays and the face design
(in particular), such as better parsing of clusters in the channel, limiting interference
in unintended directions, exposure constraints, additional losses due to face material, etc.

While one can leverage the circle's isoperimetric properties (maximizing area for
a given circumference) by using fewer antenna elements for the same desired
directivity relative to a linear or a planar array, this advantage is striking
primarily with large arrays (such as those used at the base-station end or in a
customer premises equipment or for applications such as radar, automotive,
etc.)~\cite{circ_array1}. For small-sized arrays such as those in a UE, the
complexity of designing and deploying a circular array~\cite{circ_array2}
overwhelms any potential advantages in terms of fewer antenna elements.

\section{System Setup for Spherical Coverage Studies}
\label{sec3}

\subsection{Beamforming Schemes}
\label{subsec3_a}
We study the spherical coverage CDF with four beamforming schemes that co-phase
multiple antenna transmission/reception in various ways in this work. To describe
these schemes, given a subarray of $N$ antenna elements, let ${\bf E}_{\Theta}
(\theta, \phi) = \left[ {\sf E}_{\Theta, 1}(\theta, \phi), \cdots,
{\sf E}_{\Theta, N}(\theta,\phi) \right]$ and ${\bf E}_{\Phi}(\theta, \phi) =
\left[ {\sf E}_{\Phi, 1}(\theta, \phi), \cdots, {\sf E}_{\Phi, N}(\theta,\phi)
\right]$ denote the antenna response functions in the $\Theta$ and $\Phi$
polarizations\footnote{Typically, antenna response functions are specified in the
$\Theta$ and $\Phi$ polarizations to avoid unnecessary confusion with notations such
as H- or V-polarizations that are associated with the point on the sphere where the
antenna responses are computed.}
along a certain direction\footnote{Note that
every direction in the sphere can be uniquely specified by a $(\theta, \phi)$ angle
pair. The unit-norm vector from the center of the sphere to this point is specified
as $\left[\sin(\theta)\cos(\phi) {\hspace{0.05in}} \sin(\theta) \sin(\phi) {\hspace{0.05in}}
\cos(\theta) \right]$. This classical coordinate system/transformation is also used in
3GPP channel/antenna modeling studies~\cite{3gpp_CM_rel14_38901}.} $(\theta, \phi)$
of the sphere with $\theta$ and $\phi$ denoting
the zenith/elevation and azimuth angles, respectively. As an illustration, in the scenario
of an array with $(N_{\sf x}, N_{\sf y}, N_{\sf z})$ antennas on the ${\sf X}$-${\sf Y}$-${\sf Z}$
axes where $N = N_{\sf x} N_{\sf y} N_{\sf z}$ and $\lambda/2$ inter-antenna element
spacing across all the axes, the ideal antenna response function of the $n$-th antenna
is given as~\cite{balanis}
\begin{align}
& {\hspace{-0.1in}} {\sf E}_{\Theta, n}(\theta, \phi) =
{\sf E}_{\Phi, n}(\theta, \phi) = \frac{1}{ \sqrt{ N } } \cdot
\nonumber \\
& {\hspace{0.7in}}
e^{j \hsppp \pi \cdot \left( n_{\sf x} \sin(\theta) \cos(\phi)
+ n_{\sf y} \sin(\theta) \sin(\phi) + n_{\sf z} \cos(\theta) \right)} 
, \nonumber \\
& {\hspace{2.4in}} 1 \leq n \leq N
\end{align}
where $n - 1 = n_{\sf x} + n_{\sf y} N_{\sf x} + n_{\sf z} N_{\sf x} N_{\sf y}$
with $0 \leq n_{\sf x} \leq N_{\sf x} - 1$, $0 \leq n_{\sf y} \leq N_{\sf y} - 1$
and $0 \leq n_{\sf z} \leq N_{\sf z} - 1$.

The considered beamforming schemes in this paper are as follows.
\begin{itemize}
\item {\bf \underline{Scheme 1:}} The first scheme corresponds to
MRC~\cite{tky_lo} 
in every direction $(\theta, \phi)$
of the sphere without any phase or amplitude quantization of the beamforming vector. Since
infinite-precision is assumed for phase, amplitude as well as directional resolution
(co-phasing beams are used in every direction), this scheme serves as an optimistic
upper bound on the spherical coverage performance of the UE design. In particular,
the MRC scheme over the ${\sf X}$-polarization (where ${\sf X} \in \{ \Theta,
{\hspace{0.02in}} \Phi \}$) maximizes the array gain over all possible beam weights
and this array gain in $(\theta, \phi)$ is given as
\begin{align}
G_{ {\sf mrc}, {\hspace{0.02in}} {\sf X} } (\theta, \phi) =
\max_{ \{ \alpha_i \} \hsppp : \hsppp
\sum_{i = 1}^N |\alpha_i|^2 = 1}
\left| \sum_{i = 1}^N \alpha_i^{\star}  {\hspace{0.03in}}
{\sf E}_{ {\sf X}, {\hspace{0.02in}} i}(\theta,\phi) \right|^2. 
\label{eq_MRC_objective}
\end{align}
It is straightforward to check 
that the solution to the problem in~(\ref{eq_MRC_objective}) is polarization
and~$(\theta, \phi)$-specific and is given as
\begin{eqnarray}
\alpha_i = \frac{ {\sf E}_{ {\sf X}, {\hspace{0.02in}} i}(\theta,\phi) }
{ \| {\bf E}_{ {\sf X} } (\theta, \phi) \|}
= \frac{ {\sf E}_{ {\sf X}, {\hspace{0.02in}} i}(\theta,\phi)  }
{ \sqrt{ \sum_{i = 1}^N | {\sf E}_{ {\sf X}, {\hspace{0.02in}} i}(\theta,\phi)|^2 } }
\end{eqnarray}
where $\| \cdot \|$ denotes the two-norm of a vector. We thus have
\begin{eqnarray}
G_{ {\sf mrc}, {\hspace{0.02in}} {\sf X} }(\theta, \phi) = 
\sum_{i = 1}^N | {\sf E}_{ {\sf X}, {\hspace{0.02in}} i}(\theta,\phi)|^2. 
\end{eqnarray}

\item {\bf \underline{Scheme 2:}} The second scheme considered corresponds to
EGC which is similar to the MRC scheme except that the
beamformer has an equal gain amplitude constraint for all the antennas.
It is straightforward to note that
the solution to this problem is given as
\begin{eqnarray}
G_{ {\sf egc}, {\hspace{0.02in}} {\sf X} }(\theta, \phi) =
\frac{1}{N} \left(
\sum_{i = 1}^N \left| {\sf E}_{ {\sf X}, {\hspace{0.02in}} i}(\theta,\phi) \right| \right)^2.
\end{eqnarray}
In terms of the performance comparison between MRC and EGC, we have
\begin{align}
& \sum_{i = 1}^N | {\sf E}_{ {\sf X}, {\hspace{0.02in}} i}(\theta,\phi)|^2
= G_{ {\sf mrc}, {\hspace{0.02in}} {\sf X} }(\theta, \phi)
\nonumber \\
& {\hspace{0.2in}}
\geq
G_{ {\sf egc}, {\hspace{0.02in}} {\sf X} }(\theta, \phi) =
\frac{1}{N} \left(
\sum_{i = 1}^N \left| {\sf E}_{ {\sf X}, {\hspace{0.02in}} i}(\theta,\phi) \right| \right)^2.
\end{align}
This conclusion is easy to establish and it is as intuitively expected
since the space of optimization (amplitude and phase optimization) for MRC is
bigger than that for EGC (phase-only optimization).

\item {\bf \underline{Scheme 3:}} In contrast to the above two schemes, the third
scheme is designed keeping in mind a practical implementation. This scheme corresponds
to the use of a finite-sized RF/analog beam codebook for beamforming. Note that in
the 5G-NR beam acquisition process~\cite{3gpp_CM_rel14_38912}, directional beams
providing sectoral coverage are scanned at the base-station end, while the UE
uses one fixed beam from the codebook as the base-station runs through its beams.
This process is repeated till the UE can find the best beam pair for itself (from the
codebook) and the base-station, and convey this information back to the base-station.
Thus, the worst-case beam acquisition overhead with a codebook-based scheme is
proportional to the size of the UE codebook.
To minimize the initial acquisition overhead, the UE uses a {\em pseudo-omni} beam
for each sector with a subarray of choice for this sector (P-1 phase). Beam refinement
follows by local search and optimization around the beam pair link established in the
P-1 phase (the base-station refinement is called the P-2 phase and the UE refinement
is called the P-3 phase in~\cite{3gpp_CM_rel14_38912}). 

While more beams (at both the base-station and UE ends) can result in better array
gains and hence better link performance, it comes at the cost of a higher beam
acquisition overhead. 
Therefore, for a specific UE design, a codebook size is optimally picked to tradeoff
beam acquisition overhead with link performance (see Sec.~\ref{subsec3_b} for details).
The use of the best beam from the codebook is expected to result in good spherical
coverage performance. The array gain performance over a direction $(\theta, \phi)$
with a size-$K$ codebook of beams over an $N$ antenna element subarray ($w_{ij}, \hsppp i = 1,
\cdots, N, \hsppp j = 1, \cdots, K$) is given as
\begin{align}
G_{ {\sf cbk} , {\hspace{0.02in}} {\sf X} } (\theta, \phi) = \max_{
\begin{array}{c}
w_{ij}, \hsppp j = 1, \cdots, K \\
\hsppp : \hsppp \sum_{i = 1}^N |w_{ij} |^2 = 1
\end{array} }
\left| \sum_{i = 1}^N w_{ij}^{\star}  {\hspace{0.03in}}
{\sf E}_{ {\sf X}, {\hspace{0.02in}} i}(\theta,\phi) \right|^2.
\label{eq_cbk_objective}
\end{align}
Note that as $K \rightarrow \infty$, the codebook performance can approach that
of the MRC scheme. That is, $G_{{\sf cbk}, {\hspace{0.02in}} {\sf X}}(\theta, \phi)
\rightarrow G_{{\sf mrc} , {\hspace{0.02in}} {\sf X} }(\theta, \phi)$ for every
$(\theta, \phi)$ and both polarizations. Similarly, if the codebook entries are
constrained to have equal amplitude and further if $K \rightarrow \infty$, we have
$G_{{\sf cbk}, {\hspace{0.02in}} {\sf X}}(\theta, \phi) \rightarrow
G_{{\sf egc} , {\hspace{0.02in}} {\sf X} }(\theta, \phi)$ for every $(\theta, \phi)$
and both polarizations.

\item {\bf \underline{Scheme 4:}} The fourth scheme corresponds to selecting the best
single antenna element (from amongst all the possible antennas across all the antenna
modules at the UE side) for a direction $(\theta, \phi)$ to result in:
\begin{eqnarray}
G_{ {\sf ant \hsppp sel} , {\hspace{0.02in}} {\sf X} } (\theta, \phi) = \max_{i = 1, \cdots, N}
\left|{\sf E}_{ {\sf X}, {\hspace{0.02in}} i}(\theta,\phi) \right|^2.
\end{eqnarray}
Since no beamforming array gain is realized with this scheme and the gains are
purely from antenna selection, this scheme is pessimistic in terms of the available
antenna capabilities and corresponds to a legacy beamforming solution such as
those available in prior generations (e.g., 3G or 4G) of most wireless devices.
\end{itemize}

The four schemes introduced above are studied in terms of their {\em selection
diversity} performance in this work. For this, we use the following metric:
\begin{eqnarray}
G_{ {\sf scheme} , {\hspace{0.02in}} {\sf total} } (\theta, \phi)
=
G_{ {\sf scheme} , {\hspace{0.02in}} \Theta } (\theta, \phi) +
G_{ {\sf scheme} , {\hspace{0.02in}} \Phi } (\theta, \phi)
\label{eq_metric}
\end{eqnarray}
where $G_{ {\sf scheme} , {\hspace{0.02in}} {\sf X} } (\theta, \phi)$ denotes
the array gain in the ${\sf X}$-polarization with ${\sf X} \in \{ \Theta,
{\hspace{0.02in}} \Phi \}$. In~(\ref{eq_metric}), $G_{ {\sf scheme} ,
{\hspace{0.02in}} {\sf total} } (\theta, \phi)$ captures the total array gain
seen in both polarizations in the direction $(\theta, \phi)$. This metric is used
instead of choices such as
\begin{align}
G_{ {\sf scheme} , {\hspace{0.02in}} {\sf max} } (\theta, \phi)
=
\max \left( G_{ {\sf scheme} , {\hspace{0.02in}} \Theta }  (\theta, \phi),
\hsppp G_{ {\sf scheme} , {\hspace{0.02in}} \Phi } (\theta, \phi) \right).
\label{eq_alt_metric}
\end{align}
While the metrics in~(\ref{eq_metric}) and~(\ref{eq_alt_metric}) are equivalent
(or comparable) in the boresight direction and its vicinity for a certain
subarray,~(\ref{eq_alt_metric})
can severely underestimate performance over the edge of the coverage area of the
subarray since signal strengths over both polarizations may be comparable at these
points. Over these points, gains over both polarizations may be combined with a
polarization combining scheme (such as Alamouti coding, cyclic delay diversity,
etc.) and the metric
in~(\ref{eq_metric}) appears to reflect the true selection diversity capabilities.

\begin{figure*}[htb!]
\begin{center}
\begin{tabular}{cc}
\includegraphics[height=2.1in,width=3.0in]{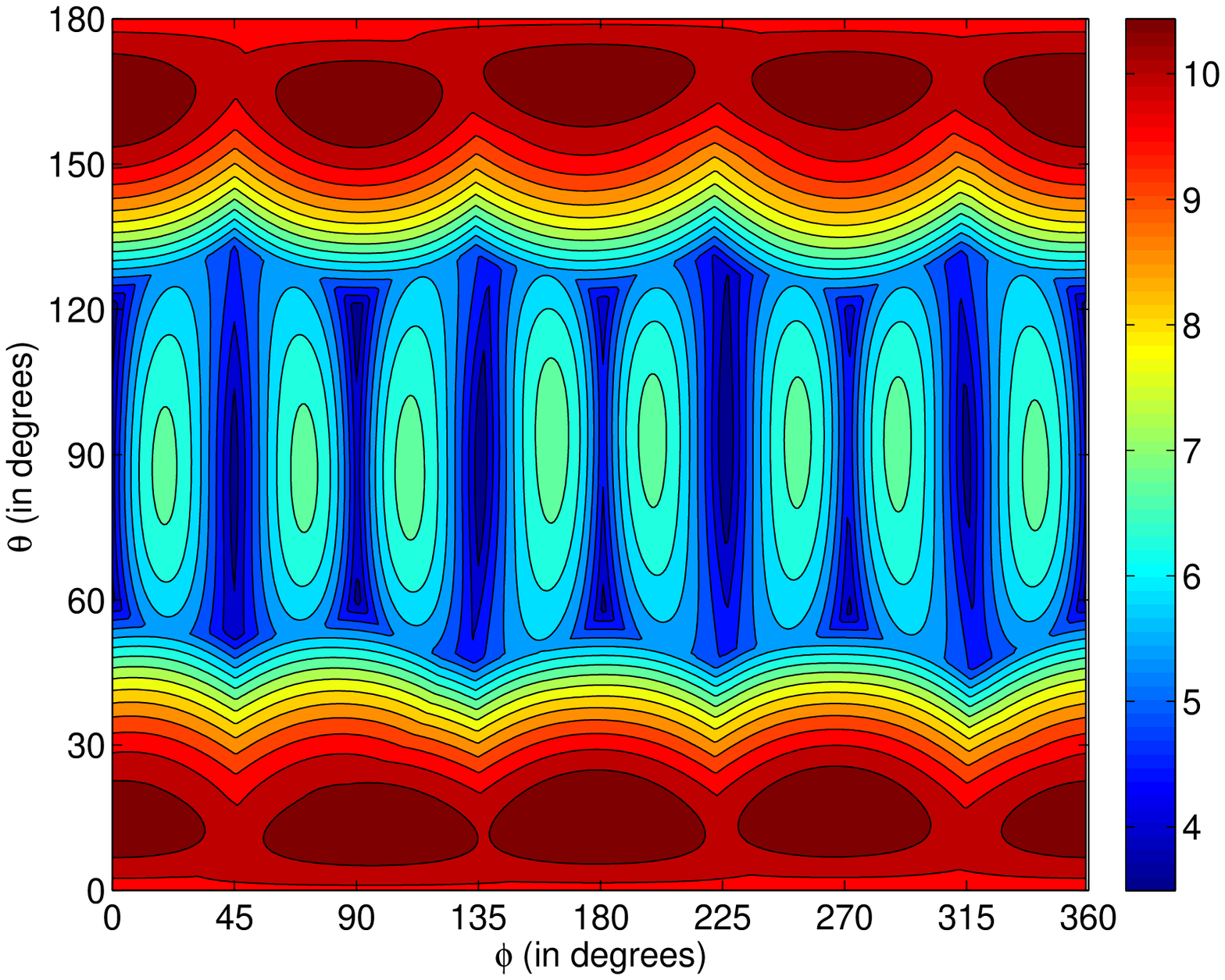}
&
\includegraphics[height=2.1in,width=3.0in]{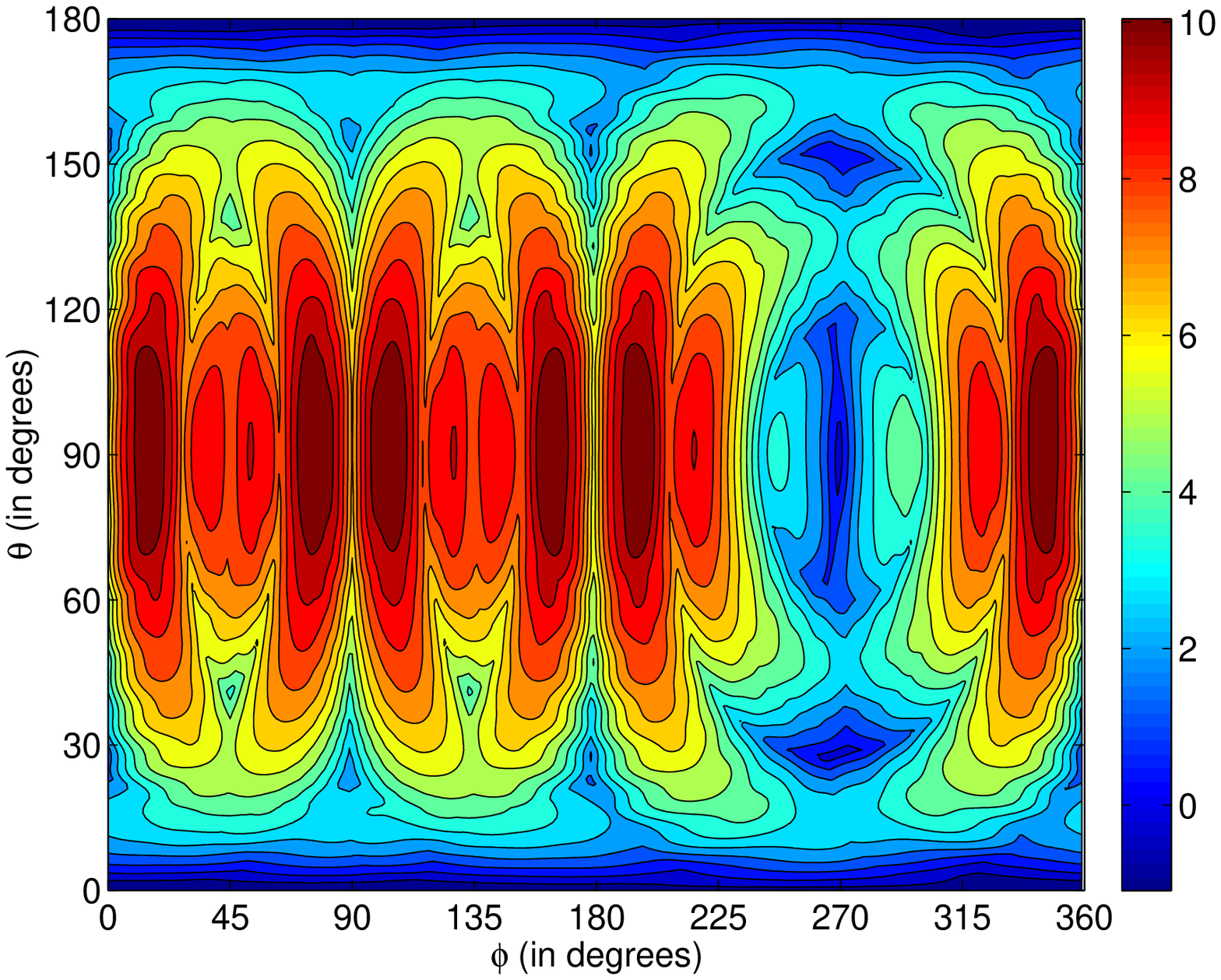}
\\
(a) & (b)
\end{tabular}
\caption{\label{fig_codebooks}
View of codebooks' array gain performance over the sphere (represented in the
$\theta$-$\phi$ Cartesian plane) for the (a) face and (b) edge designs.}
\end{center}
\end{figure*}

\subsection{RF/Analog Beam Codebook Design}
\label{subsec3_b}
As noted in Sec.~\ref{subsec3_a}, with hierarchical beamforming as in 5G-NR, the P-1 phase is
typically\footnote{Note that the initial beam acquisition can also be performed
over periodically configured channel state information reference signals (CSI-RS).}
performed over a burst set of sychronization signal blocks (SSBs)~\cite{survey_5gnr_ericsson}.
Beamformed transmissions over different beams (up to $64$ beams are allowed in 5G-NR for
millimeter wave frequencies) can be used over multiple SSBs in a burst set of $5$ ms duration
and the burst set can be repeated with periodicity that is one of either $5$ ms,
$10$ ms, or $20$ ms. With initial practical implementations as well as with the
initial acquisition phase of UEs, it is expected that the SSB burst set periodicity is
set to $20$ ms (which is assumed in this work). On the other
hand, P-2 and P-3 beam refinements are typically performed with aperiodic CSI-RS
symbols in a UE-specific manner. In some possibilities with low-cost and
low-complexity base-stations, P-3 beam refinement can also be performed over
SSB signals.

We assume that both the base-station and all the UE designs considered in this work are
powered by two RF chains, where each RF chain is excited by one (orthogonal) polarization.
This is a reasonable assumption for the initial generation of millimeter wave systems.
That is, independent beam weights can be set for either of the two RF chains corresponding
to different subarrays excited by these RF chains. Thus, two subarrays at the UE side can
be beam trained by the base-station at the same time.
With UE side beam switching constraints in mind,
we assume that the UE uses a single {\em pseudo-omni} beam per subarray over each SSB
burst set. Further, with UE side power-performance tradeoffs in mind, a single
antenna exciting {\em pseudo-omni} beam is used from all the subarrays for the P-1 phase.
While more complicated {\em pseudo-omni} beam choices can be used, the considered design
is representative of practical implementations. Thus, for the face design with $8$
subarrays over two polarizations, the initial beam acquisition overhead in the P-1
phase is $20$ ms $\times 8$ subarrays/$2$ polarizations, which equals $80$ ms. Similarly,
for the edge design with $6$ subarrays, the initial beam acquisition overhead corresponds
to $60$ ms. These numbers appear
to be representative of the initial beam acquisition overheads expected in inter-operability
development and testing trials, and commercial operations of 5G-NR.


In terms of peak performance, in order to compare the UE designs introduced in
Sec.~\ref{subsec2_a}, it is important to perform a {\em fair} comparison of the
codebook-based beamforming scheme across these designs. With different codebook sizes,
the beam acquisition latencies can be different. To address these concerns, for the
face design, $4$
beams are used for each polarization of the $2 \times 2$ dual-polarized patch subarrays,
and $2$ beams are used for each dipole subarray leading to $12$ beams per antenna module
as well as $12$ beams per polarization. For the edge design, $4$ beams are used for each
polarization of the $4 \times 1$ dual-polarized patch subarray for $8$ beams per module and
$12$ beams per polarization. Since the codebook sizes are $24$ for both the designs, the
performance of these two UE designs can be compared fairly. Note that while more complicated
and different-sized codebooks can be considered for the two designs and their performance
can be compared with some performance penalty function (e.g., a $3$ dB penalty for a doubling
of the codebook in one design relative to the other, etc.), the method proposed here is
reasonable for practical implementations. 

If UE-specific CSI-RS symbols are used for beam refinement, since multiple symbols can be
configured for CSI-RS in a downlink-specific subframe, P-3 beam refinement can be performed
within $1$-$2$ slots even under the assumption of multi-symbol averaging for signal-to-noise
ratio (SNR) enhancement.
As an illustration, since each subarray has at most $4$ refined/narrow beams for a {\em pseudo-omni}
beam, assuming a four symbol averaging, we would need at most $16$ symbols which can be
accommodated within $2$ slots since $14$ symbols make a slot. Under a $60$ kHz subcarrier
spacing for millimeter wave transmissions, the worst-case beam refinement overhead is thus
less than $0.50$ ms (which is significantly smaller than that accrued in the P-1 phase).
Alternately, if SSBs at $20$ ms periodicity are used for P-3 beam refinement, the worst-case
beam refinement overhead for the P-3 phase (assuming $4$ narrow beams) is $20$ ms
$\times 4$ beams/$1$ polarization, which equals an additional $80$ ms time-period. Note that
in this calculation, beam refinement is constrained to the RF chain corresponding to the
selected subarray (unlike the initial beam acquisition phase over two RF chains).

The individual
beam weights in the codebook can be optimally designed to cover certain angular regions over
the sphere. While the beam design process in itself can be implementation-specific or proprietary,
general design principles are expounded in~\cite{raghavan_jstsp}. In this work, the beams for
the $4 \times 1$ subarrays are designed such that each beam results in a beamwidth of
$\approx 30^{\sf o}$. Similarly, the beams for the $2 \times 2$ and $2 \times 1$ subarrays
have a beamwidth of $\approx 55^{\sf o}$. All the beam weights (for either design)
are constrained to meet an equal amplitude and a five-bit phase shifter resolution.
Fig.~\ref{fig_codebooks} presents the array gain performance with the codebooks over the sphere
(represented on a $\theta$-$\phi$ Cartesian plane) for the face and edge designs. For each point
$(\theta, \phi)$ over the sphere, the best representative from each design's codebook is used
here. Clearly, from Fig.~\ref{fig_codebooks}, we observe that the codebooks are designed to meet
good array gain performance over a significant fraction of the sphere.

\subsection{Blockage Modeling}
\label{subsec3_c}
Hand and body blockage can be substantial at millimeter wave carrier frequencies
(relative to sub-$6$ GHz frequencies) since the size of many physically small
objects in the proximity of the antennas become electrically comparable with the
wavelength of propagation.

In the 3GPP channel modeling document TR 38.901, a blockage model is
proposed~\cite[pp.\ 53-57]{3gpp_CM_rel14_38901} to capture these detrimental
effects under two variants: a stochastic variant (Option A) and a map-based/ray
tracing-based variant
(Option B). The stochastic variant proposes a spherical coverage blockage
tailored to the hand in Portrait or Landscape orientations around a UE modeled
to form factor considerations. As illustrated in Table~\ref{table_hand_blockage},
this model (labeled ``Model 1'' in this work) is captured by the center of the
blocker ($\phi_1$, $\theta_1$), and the angular spread of the blocker ($x_1$,
$y_1$) in azimuth and elevation with the blocking angles captured as $\phi
\in \left[ \phi_1 - \frac{x_1}{2}, \hspp \phi_1 + \frac{x_1}{2} \right]$ and
$\theta \in \left[ \theta_1 - \frac{y_1}{2}, \hspp \theta_1 + \frac{y_1}{2}
\right]$ in azimuth and elevation, respectively. Over this spatial region, a
simplistic flat $30$ dB loss is assumed.

It is understood that the 3GPP blockage model is quite
pessimistic~\cite{vasanth_blockage_tap2018} relative to form factor UE designs
due to the use of horn antenna measurements (with smaller beamwidths) to model
hand blockage loss. In this context, based on $28$ GHz experimental prototype
studies~\cite{vasanth_comm_mag_16}, a modified blockage model (labeled as
``Model 2'' in this work) is proposed in~\cite{vasanth_blockage_tap2018}. In
this model, the spatial blockage region is retained from the 3GPP model and a
log-normal blockage loss term, as summarized in Table~\ref{table_hand_blockage},
is used. We study the spherical coverage CDFs with these two blockage models in
this work.

\begin{table*}[htb!]
\caption{Hand Blockage Models Considered in This Work}
\label{table_hand_blockage}
\begin{center}
\begin{tabular}{|c||c|c|c|c|c|}
\hline
Scenario & $\phi_1$ & $x_1$ & $\theta_1$ & $y_1$ & Blockage loss (in dB) \\
\hline
\hline
Portrait mode & $260^{\sf o}$ & $120^{\sf o}$ & $100^{\sf o}$ &
$80^{\sf o}$ & 
Model 1 \hspp : \hspp $30 \hspp {\sf dB}$
\\
\cline{1-5}
Landscape mode & $40^{\sf o}$ & $160^{\sf o}$ & $110^{\sf o}$ &
$75^{\sf o}$ & 
Model 2 \hspp : \hspp ${\cal N}(\mu = 15.3 \hspp {\sf dB}, \hspp \sigma = 3.8 \hspp {\sf dB})$
\\
\hline
\hline
\end{tabular}
\end{center}
\end{table*}

At this stage, it is to be noted that the dielectric properties of the
skin tissue (such as the relative dielectric constant and conductivity) determine the
penetration depth of the electromagnetic radiation into the hand and its reflection. At
$28$ GHz, it is observed that the penetration depth into the hand is very small and a
significant fraction of the energy is reflected. The hand blockage loss is a function
of the shape of the hand (its curvature, roughness of skin tissue, etc.) and the model
presented in~\cite{vasanth_blockage_tap2018} and used in this work is reflective of
these ensemble trends. That said, the precise reflection response and penetration loss of
different materials to millimeter wave frequencies is a function of the material, incidence
angle and polarization~\cite[Sec.~IV-A and Fig.~5]{vasanth_tap2018}. While similar
overall behaviors are seen with the hand to the two polarizations, more work is
necessary to understand
the precise differences, if any. Thus, the hand blockage model used here can be seen to be
a good first effort at understanding the impact of hand impairments for UE design.

\section{Spherical Coverage CDF Tradeoffs} 
\label{sec4}

We now present results on spherical coverage CDF tradeoffs with the two
UE designs considered in this work. For this, as explained in Sec.~\ref{sec3},
individual antenna element response functions in both polarizations are computed
using the HFSS commercial software suite~\cite{hfss} with a $1^{\sf o}$ precision
in azimuth and elevation. Beamforming gains are computed with
MRC, EGC, an RF/analog beam codebook and single antenna selection with the
diversity performance metric as described in Sec.~\ref{subsec3_a}. Spherical
coverage CDF is computed as described in Appendix~\ref{app_spherical}.

\subsection{Freespace Performance}
\begin{figure*}[htb!]
\begin{center}
\begin{tabular}{cc}
\includegraphics[height=2.3in,width=3.0in]{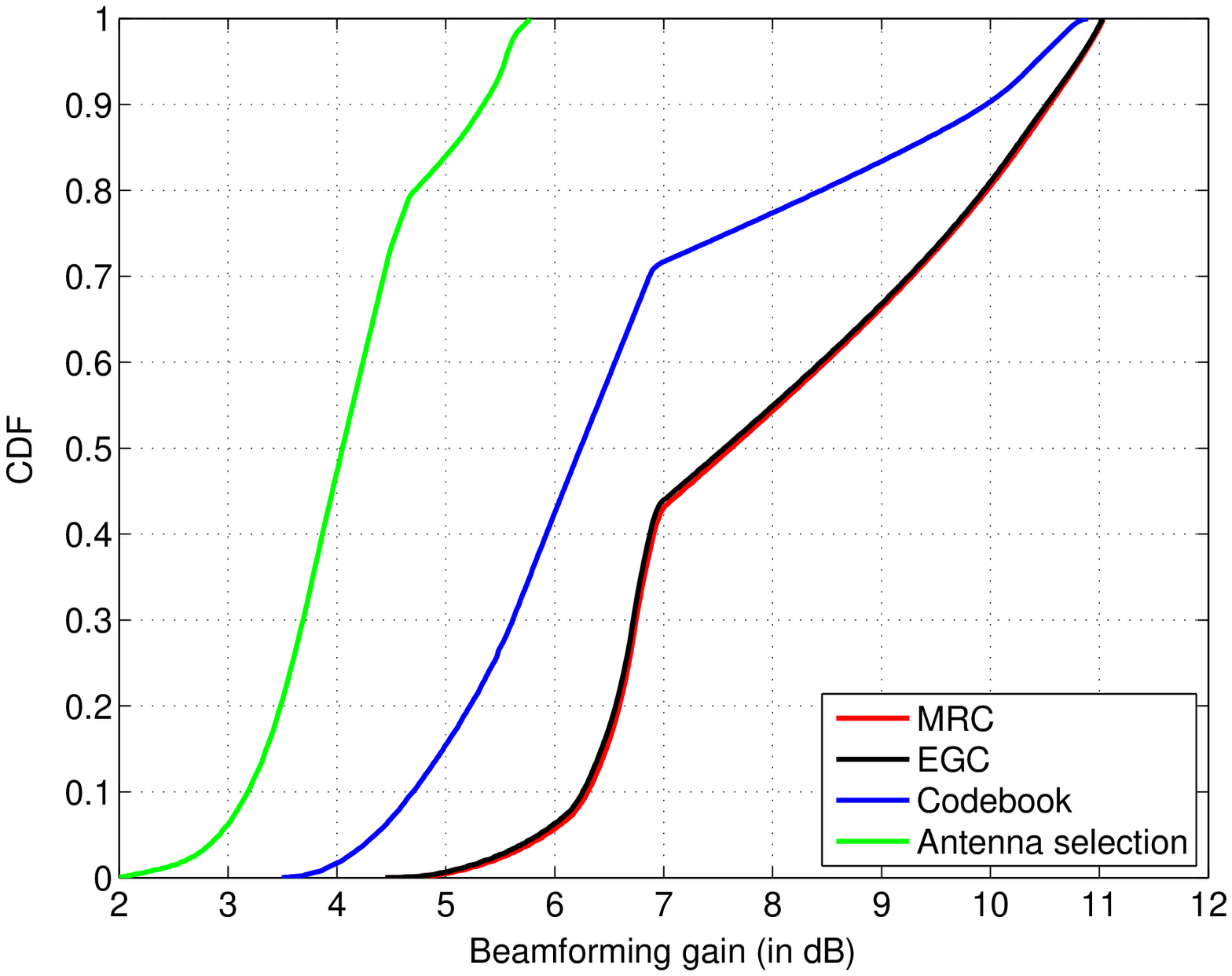}
&
\includegraphics[height=2.3in,width=3.0in]{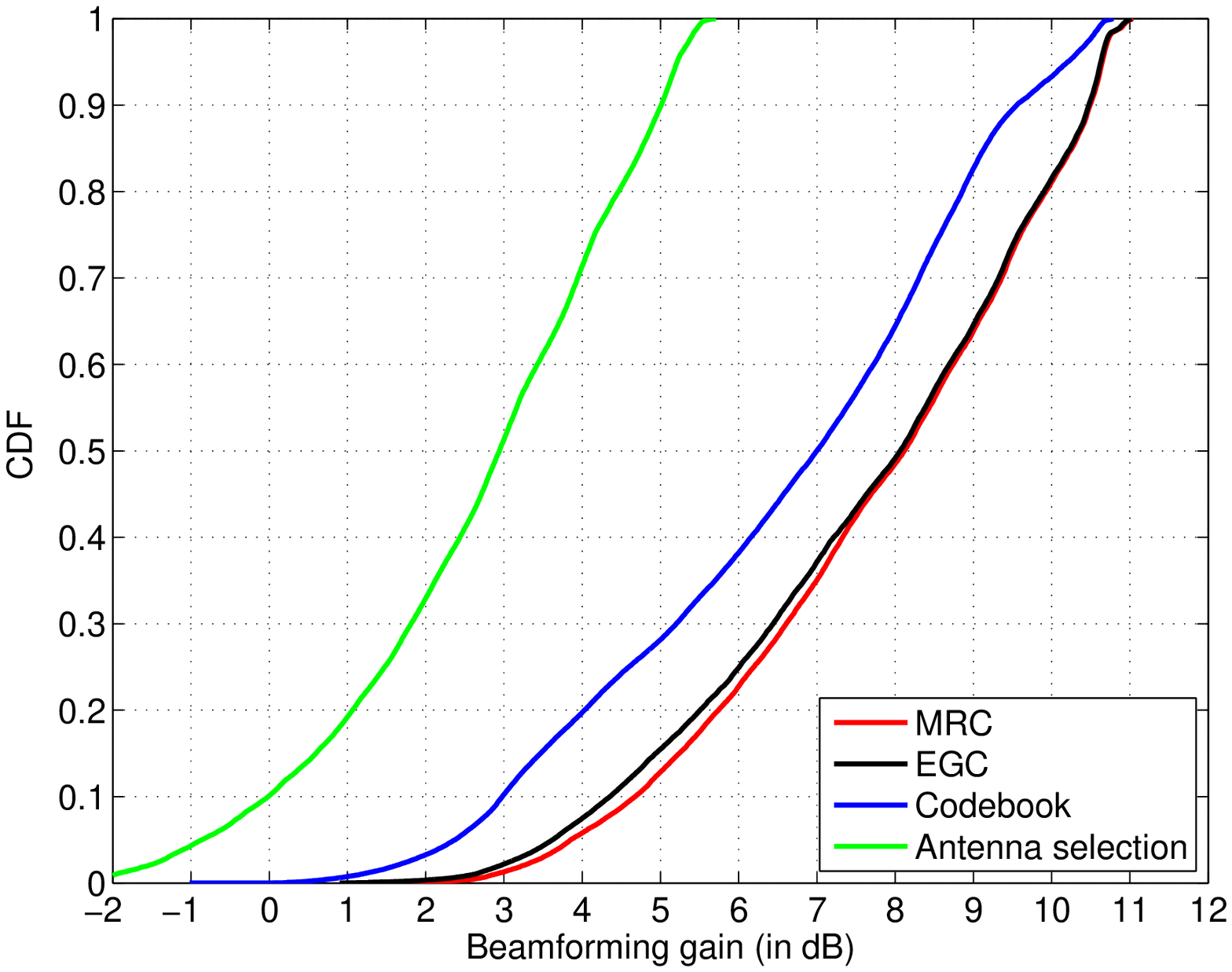}
\\
(a) & (b)
\end{tabular}
\caption{\label{fig_bf_Freespace}
Array gain performance of different beamforming schemes in Freespace for the (a) face
and (b) edge designs.}
\end{center}
\end{figure*}

In our first study, in
Figs.~\ref{fig_bf_Freespace}(a)-(b), we describe the beamforming array gain tradeoffs
for the UE designs with these four schemes in Freespace (that is, with no hand blockage).
With both the designs, we observe that the EGC scheme performs as well as the MRC scheme
over the entire sphere. This conclusion implies that phase-only control is sufficient to
obtain the optimal
spherical coverage and the cost associated with amplitude control can be forsaken
with minimal performance penalties. This also motivates the design of RF/analog beam
codebooks with only phase shifter control as done in Sec.~\ref{subsec3_b}. This
conclusion stems from the fact that
all the antennas that make a certain subarray have similar/comparable amplitudes over
the whole sphere and no specific antenna sees an anomalous behavior (relative to others)
necessitating amplitude control.

For both the UE designs, the RF/analog beam codebooks are within $1$-$2$ dB of the
MRC/EGC performance suggesting the goodness of the codebook design principles. However,
the worst-case points of the codebook's performance 
are $7$ dB and $10$ dB away from the peak gain for the face and edge designs,
respectively. While this observation could suggest that there are significant gaps
relative to MRC/EGC performance, this is a na\"ive conclusion that needs to be tested
with real impairments. We will see subsequently that both the face and edge designs are
competitive with hand blockage.

With the edge design, single antenna selection is approximately $5$-$6$ dB worse than
MRC/EGC. This gap can be explained as the co-phasing gain from four antenna subarrays
used in this design. On the other hand, with the face design, this gap reduces from $6$ dB
at the peak to $3$ dB at the tail corresponding to the switch from a $2\times 2$ patch
subarray to a $2 \times 1$ dipole subarray. In terms of codebook performance relative
to MRC/EGC, the edge design shows a near-constant gap over the CDF curve ($\approx 1$ dB).
On the other hand, the face design appears to have a gap that increases from the peak to
the tail. This can be attributed to: i) loss in array gain as we move from the beams'
boresight steering direction to the edge of coverage of each beam, and ii) switch from a
four element subarray to a two element subarray. From a pictorial view of the codebooks
in Fig.~\ref{fig_codebooks}, we observe that the edge design has coverage holes
only/mostly over the poles (which are discounted with the $\sin(\theta)$ factor in
the spherical coverage computation --- See Appendix~\ref{app_spherical}), whereas
the face design has coverage holes at random points over the sphere accounting for
the degradation in codebook performance from the peak to the tail.

\subsection{Performance of the Face Design with Hand Blockage}
\begin{figure*}[htb!]
\begin{center}
\begin{tabular}{cc}
\includegraphics[height=2.3in,width=3.0in]{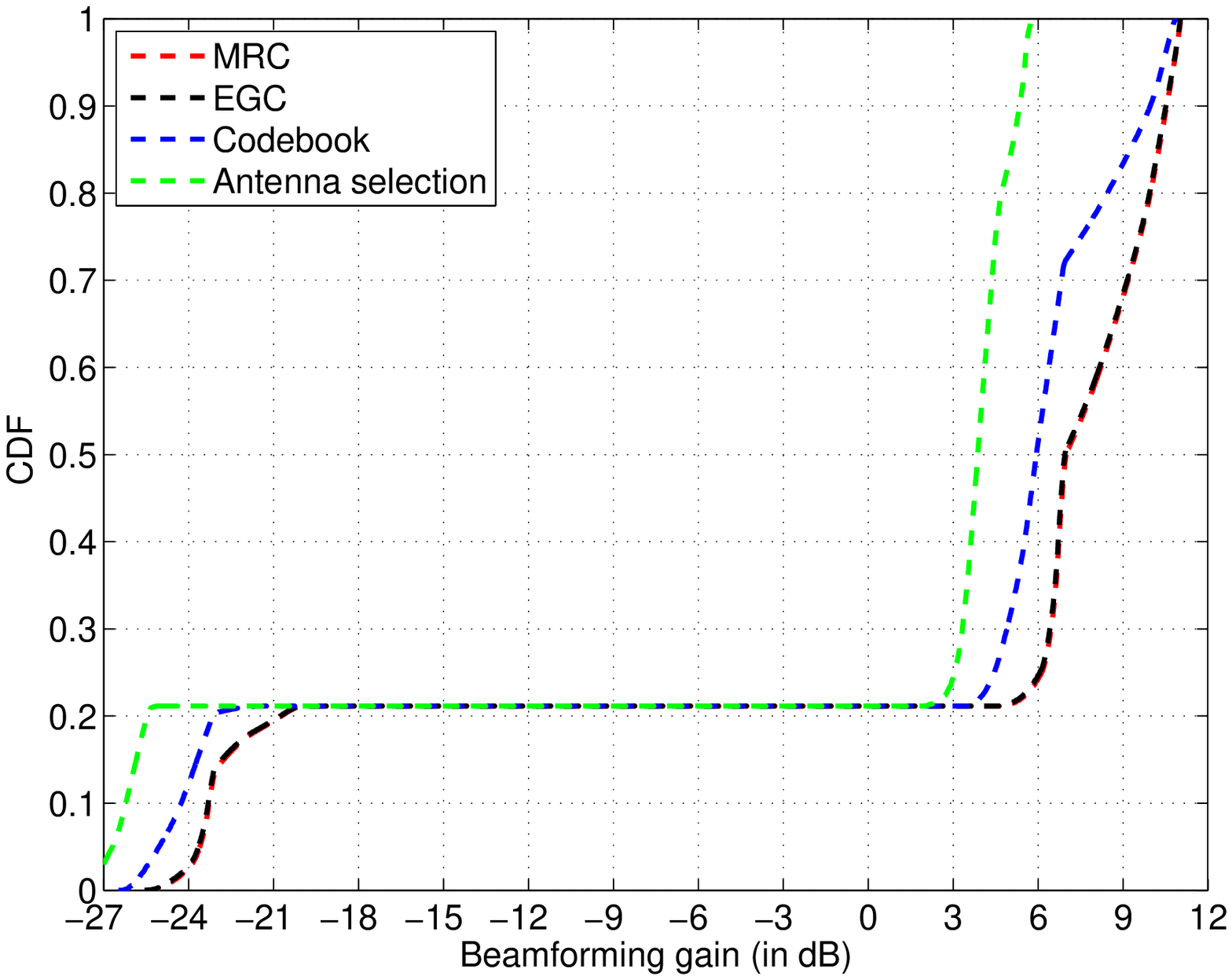}
&
\includegraphics[height=2.3in,width=3.0in]{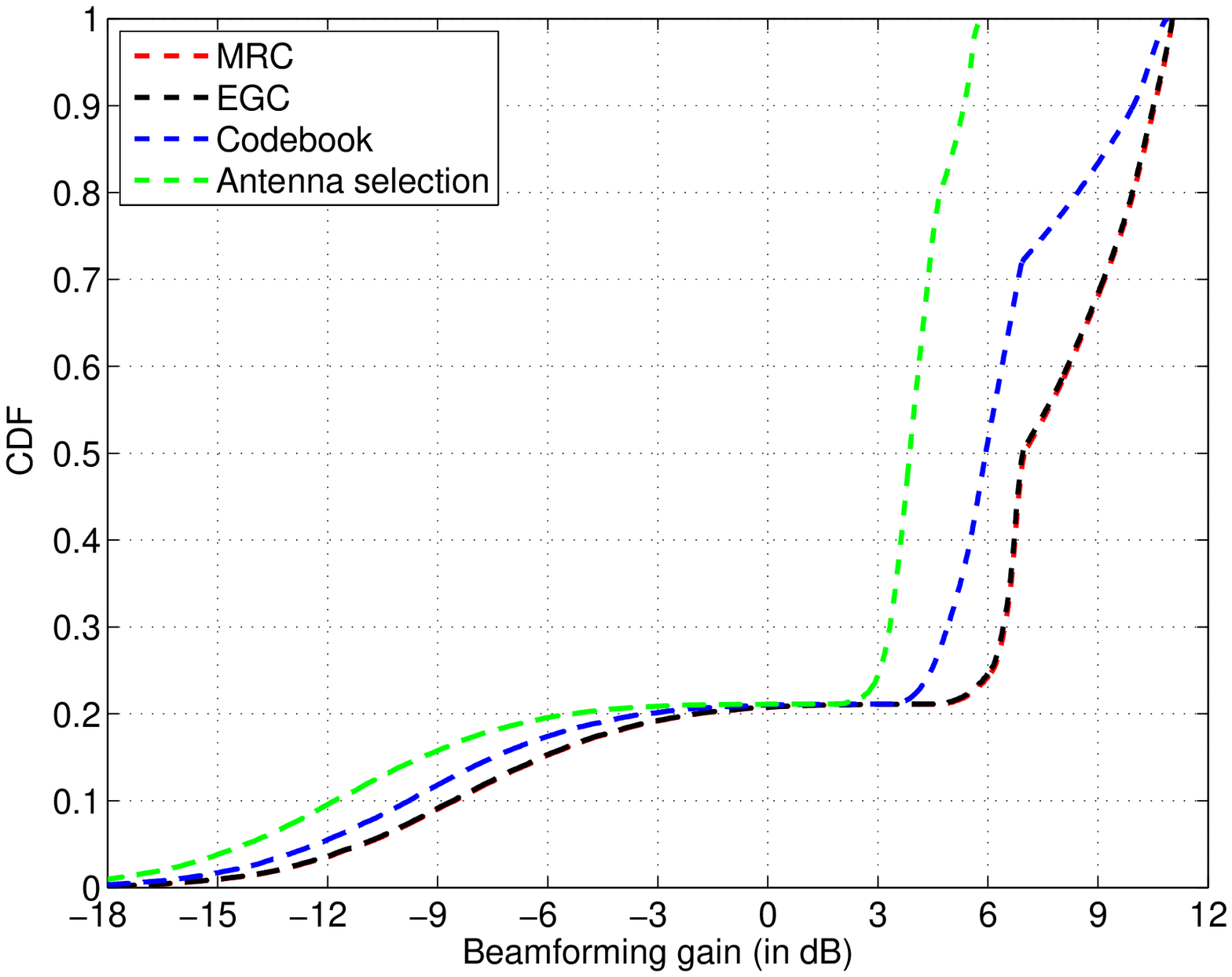}
\\
(a) & (b) \\
\includegraphics[height=2.3in,width=3.0in]{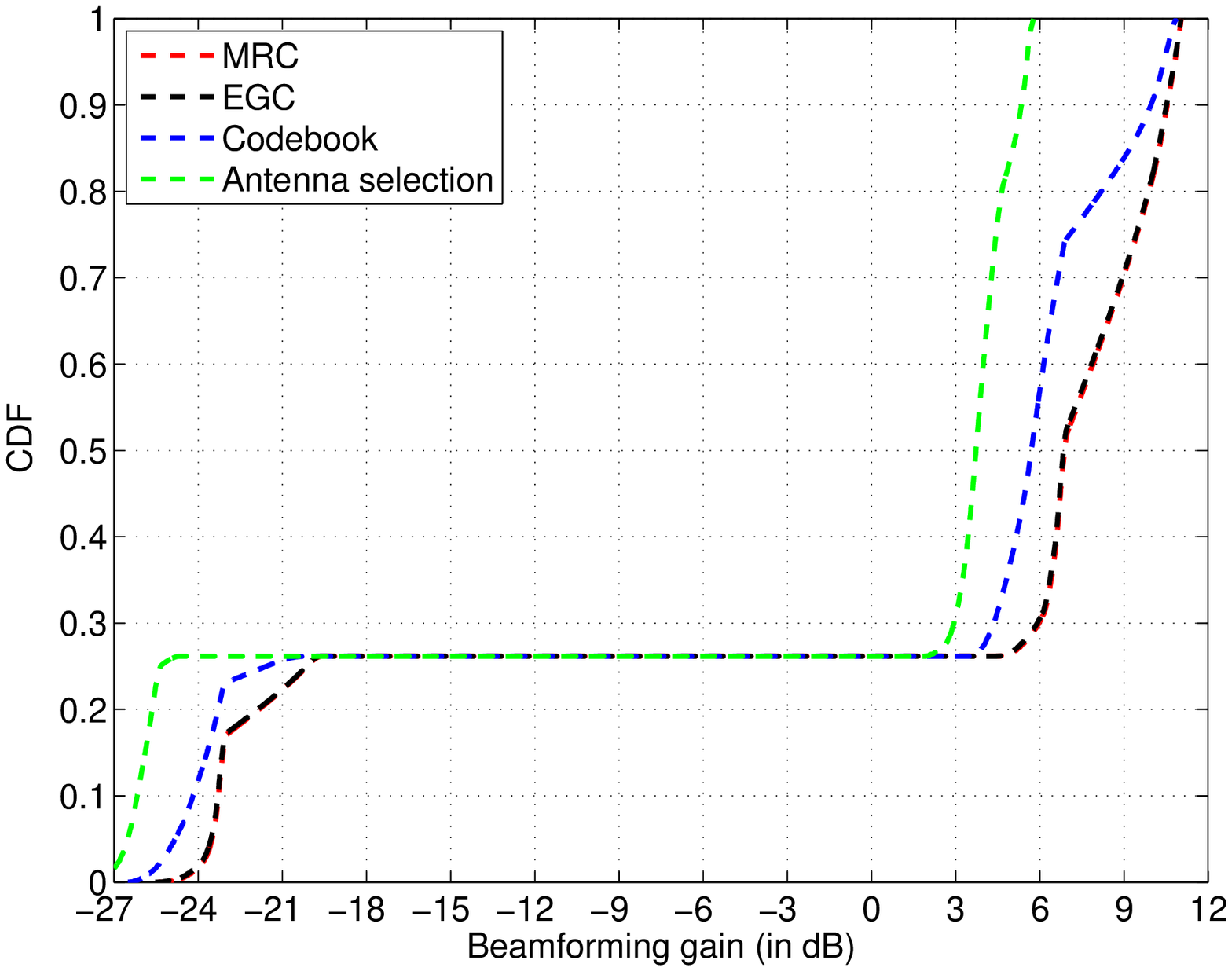}
&
\includegraphics[height=2.3in,width=3.0in]{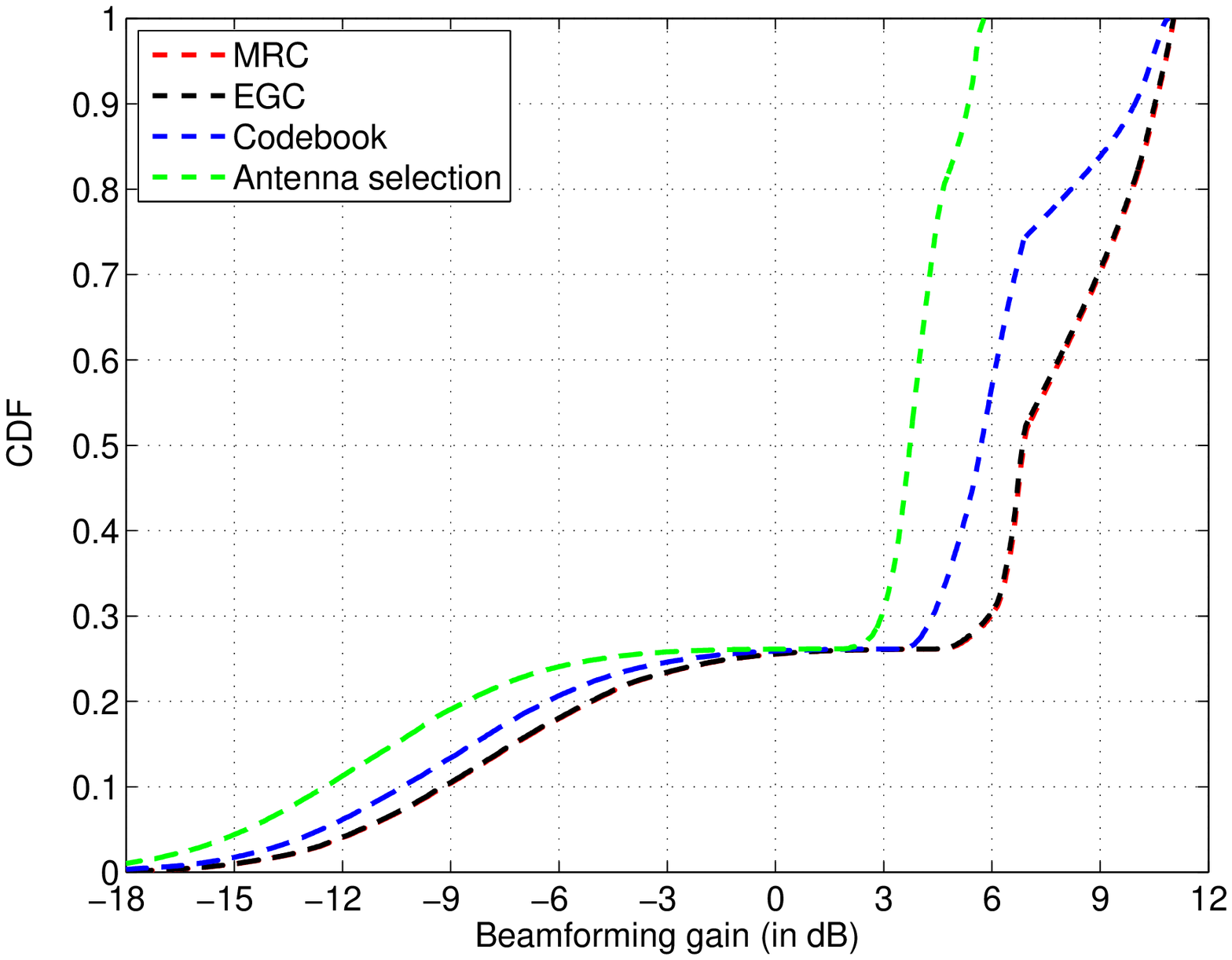}
\\ {\hspace{0.2in}}
(c) & 
(d)
\end{tabular}
\caption{\label{fig_Portrait_Landscape_Design1}
Array gain performance of the face design with blockage model in (a)-(b) Portrait mode
(Models 1 and 2) and (c)-(d) Landscape mode (Models 1 and 2).}
\end{center}
\end{figure*}

We now study
the performance of the face design with the blockage models described in
Sec.~\ref{subsec3_c}. Figs.~\ref{fig_Portrait_Landscape_Design1}(a)-(b) present the
performance of the different schemes with the two blockage models in Portrait mode,
whereas Figs.~\ref{fig_Portrait_Landscape_Design1}(c)-(d) present the performance
with the two models in Landscape mode.

From Table~\ref{table_hand_blockage}, the blockage region in Portrait and Landscape
modes occupy the following fraction of physical/spatial angles: 
\begin{align}
& {\sf Physical} \hspp {\sf angle} \hspp {\sf loss} \Big|_{\sf Portrait}
= \frac{ 120^{\sf o} \times
80^{\sf o}}{ 360^{\sf o} \times 180^{\sf o}}
= 14.81\% \\
& {\sf Physical} \hspp {\sf angle} \hspp {\sf loss} \Big|_{\sf Landscape}
= \frac{ 160^{\sf o} \times
75^{\sf o}}{ 360^{\sf o} \times 180^{\sf o}}
= 18.52\%,
\end{align}
respectively. Since the spatial angles need to be weighted based on the Jacobian
(see Appendix~\ref{app_spherical}), these blocked angles correspond to a CDF loss of
\begin{align}
& {\sf CDF} \hspp {\sf loss} \Big|_{\sf Portrait}
\nonumber \\
& = \frac{1}{4\pi}
\int_{\phi = \phi_{ {\sf p}, \hsppp {\sf l}} }^{ \phi_{ {\sf p}, \hsppp {\sf u}} }
\int_{\theta = \theta_{ {\sf p}, \hsppp {\sf l}} }^{ \theta_{ {\sf p}, \hsppp {\sf u}} }
\sin(\theta) \cdot d\theta d\phi = 21.07\%
\label{eq_CDF_loss_Portrait}
\\
& {\sf CDF} \hspp {\sf loss} \Big|_{\sf Landscape}
\nonumber \\
& = \frac{1}{4\pi}
\int_{\phi = \phi_{ {\sf l}, \hsppp {\sf l}} }^{ \phi_{ {\sf l}, \hsppp {\sf u}} }
\int_{\theta = \theta_{ {\sf l}, \hsppp {\sf l}} }^{ \theta_{ {\sf l}, \hsppp {\sf u}} }
\sin(\theta) \cdot d\theta d\phi = 26.00\%,
\label{eq_CDF_loss_Landscape}
\end{align}
where $\phi_{ {\sf p}, \hsppp {\sf l}} = 200^{\sf o} \cdot \pi/180$,
$\phi_{ {\sf p}, \hsppp {\sf u}} = 320^{\sf o} \cdot \pi/180$,
$\theta_{ {\sf p}, \hsppp {\sf l}} = 60^{\sf o} \cdot \pi/180$,
$\theta_{ {\sf p}, \hsppp {\sf u}} = 140^{\sf o} \cdot \pi/180$, and
$\phi_{ {\sf l}, \hsppp {\sf l}} = -40^{\sf o} \cdot \pi/180$,
$\phi_{ {\sf l}, \hsppp {\sf u}} = 120^{\sf o} \cdot \pi/180$,
$\theta_{ {\sf l}, \hsppp {\sf l}} = 72.5^{\sf o} \cdot \pi/180$,
$\theta_{ {\sf l}, \hsppp {\sf u}} = 147.5^{\sf o} \cdot \pi/180$.
The performance degradation in the tails of the Portrait and Landscape modes
correspond to the CDF loss region estimates in~(\ref{eq_CDF_loss_Portrait})
and~(\ref{eq_CDF_loss_Landscape}), as expected.
The major difference between the two sets of curves is that the flat $30$ dB
loss assumed with the 3GPP model (Model 1) renders the tail region completely
irretrievable and the performance loss over this region is abrupt/dramatic.
On the other
hand, with a log-normal loss model (Model 2), this loss in performance is
smoother allowing for some recovery over certain directions. Nevertheless, in
general, it appears that blockage leads to a {\em bimodal} behavior of almost no
loss over the unblocked region and essentially irretrievable loss over the blocked region.

\subsection{Freespace vs.\ Blockage}
\begin{figure*}[htb!]
\begin{center}
\begin{tabular}{cc}
\includegraphics[height=2.3in,width=3.0in]{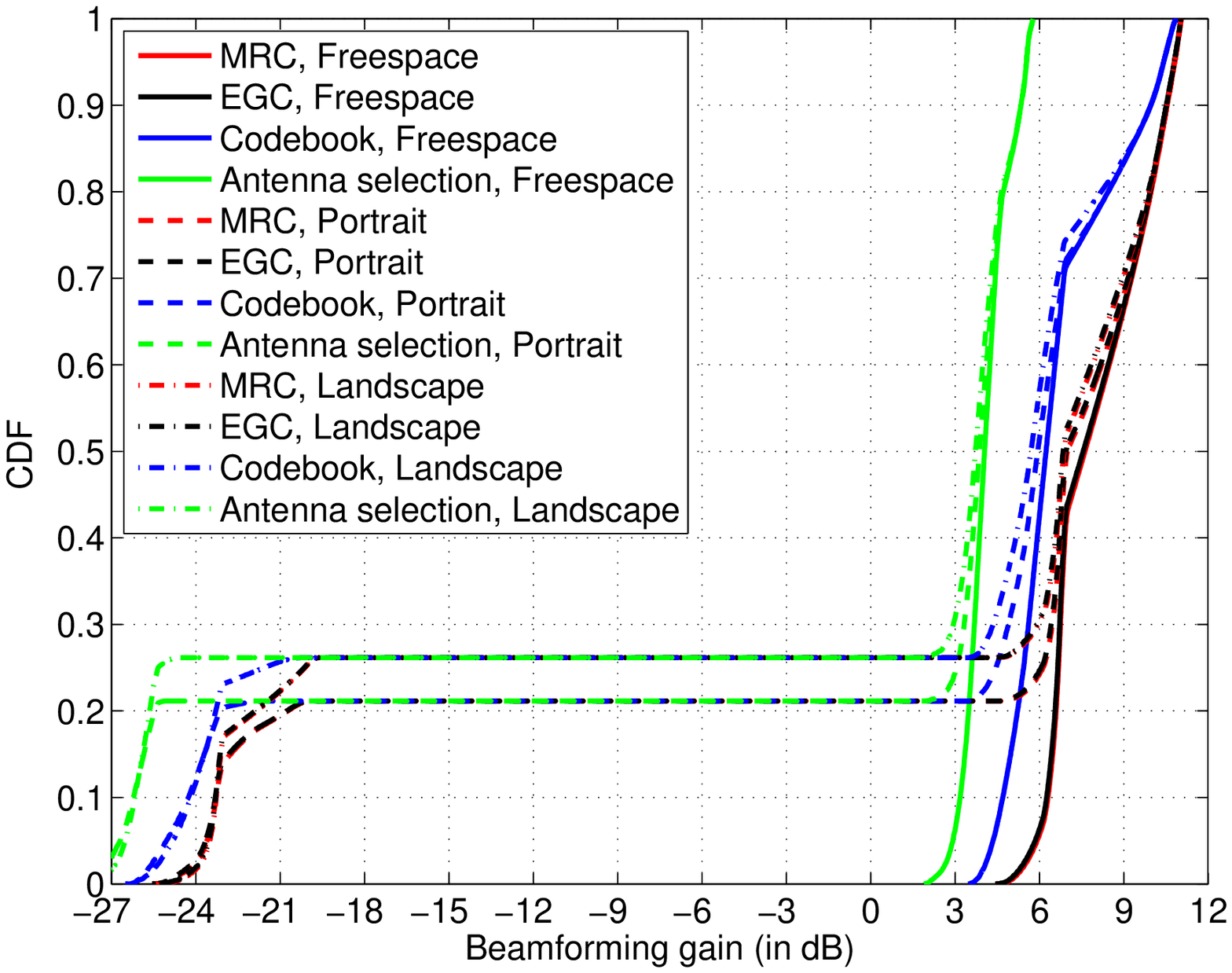}
&
\includegraphics[height=2.3in,width=3.0in]{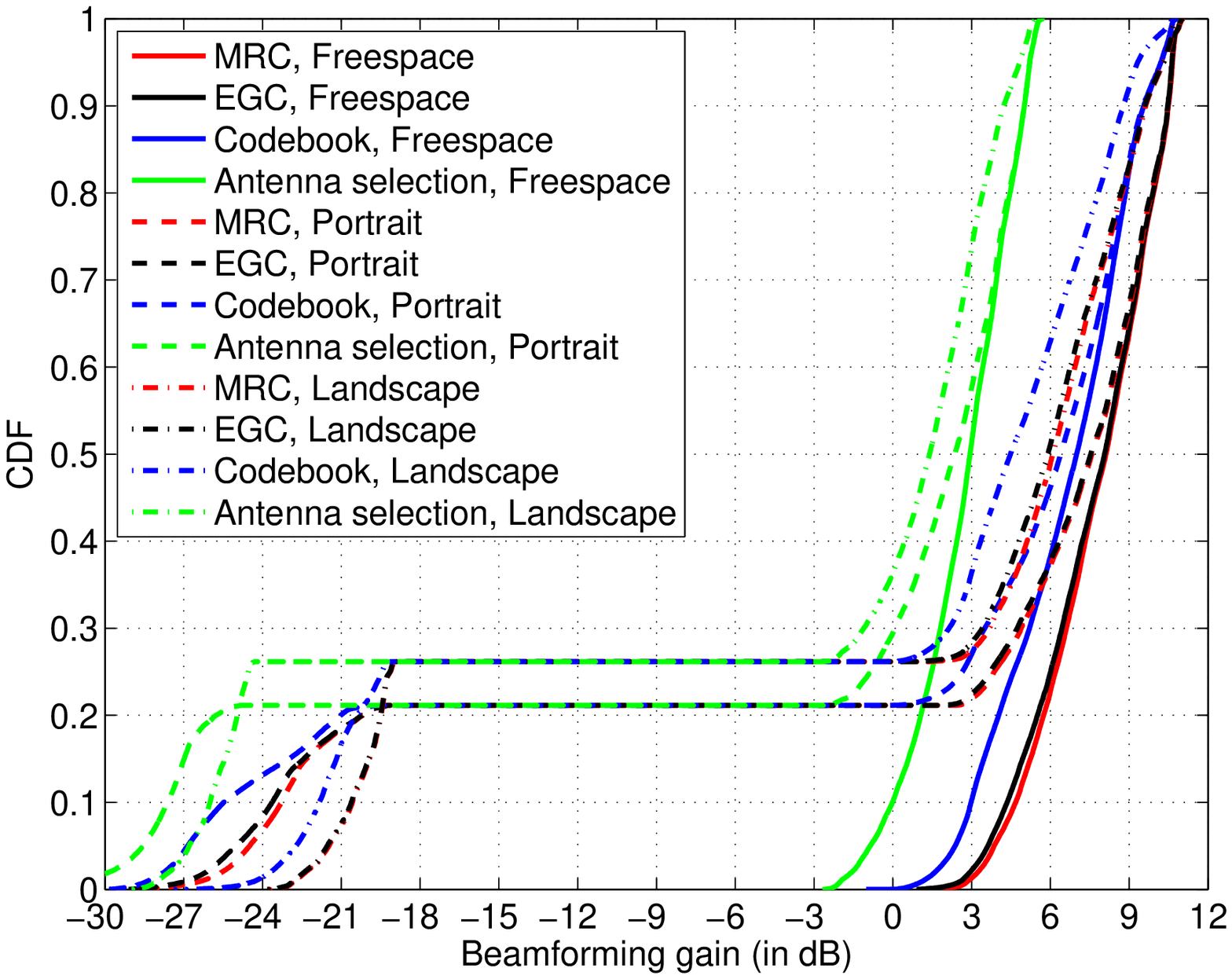}
\\
(a) & (b)
\end{tabular}
\caption{\label{fig_all_Model1}
Array gain performance with Freespace and blockage in Portrait and Landscape (Model 1) for
the (a) face and (b) edge designs.}
\end{center}
\end{figure*}

\begin{figure*}[htb!]
\begin{center}
\begin{tabular}{cc}
\includegraphics[height=2.3in,width=3.0in]{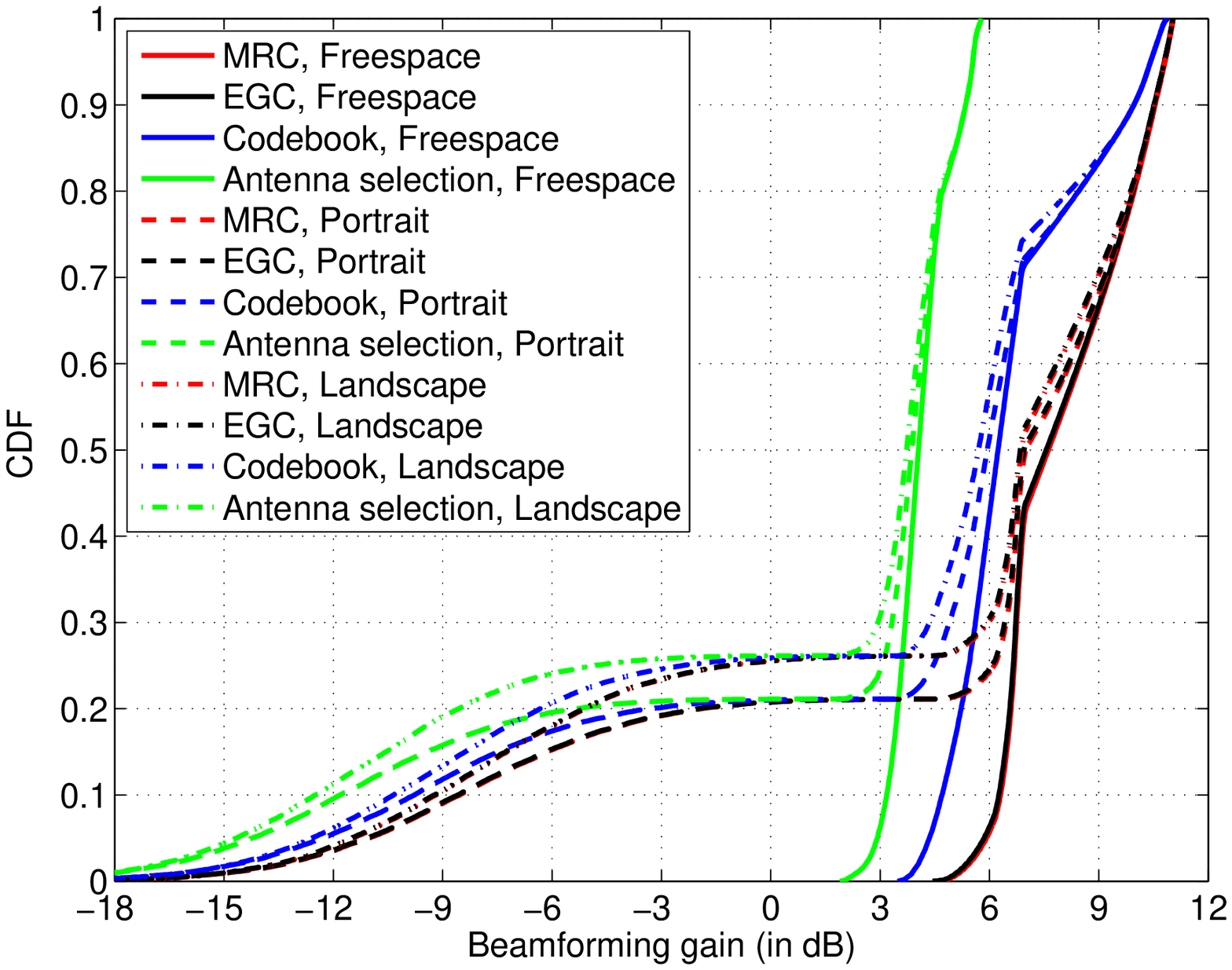}
&
\includegraphics[height=2.3in,width=3.0in]{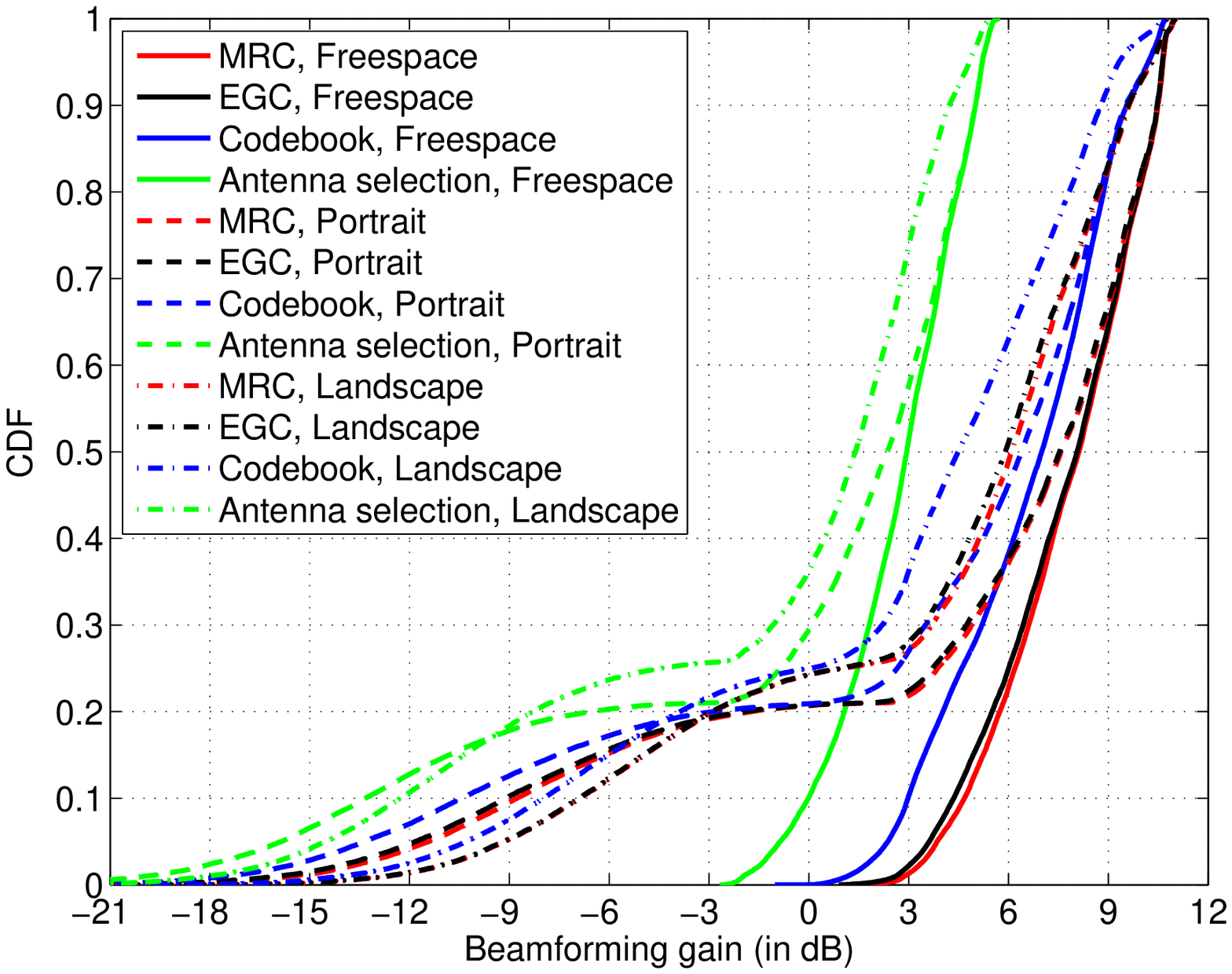}
\\
(a) & (b)
\end{tabular}
\caption{\label{fig_all_Model2}
Array gain performance with Freespace and blockage in Portrait and Landscape (Model 2) for
the (a) face and (b) edge designs.}
\end{center}
\end{figure*}

Figs.~\ref{fig_all_Model1}
and~\ref{fig_all_Model2} present the comparison between Freespace performance
for the two designs and hand blockage in Portrait and Landscape modes with
Models 1 and 2, respectively. All the four beamforming schemes are considered in
these plots. Fig.~\ref{fig_all_Model1} reinforces the earlier finding of the
blocked region being completely irretrievable, independent of which UE design is used.
On the other hand, Fig.~\ref{fig_all_Model2} shows a smoother degradation over
the blocked angles with both the designs and with the precise set/quantum of angles
recoverable with blockage depending on the UE design.

\subsection{Comparisons Across Designs}
\begin{figure*}[htb!]
\begin{center}
\begin{tabular}{cc}
\includegraphics[height=2.3in,width=3.0in]{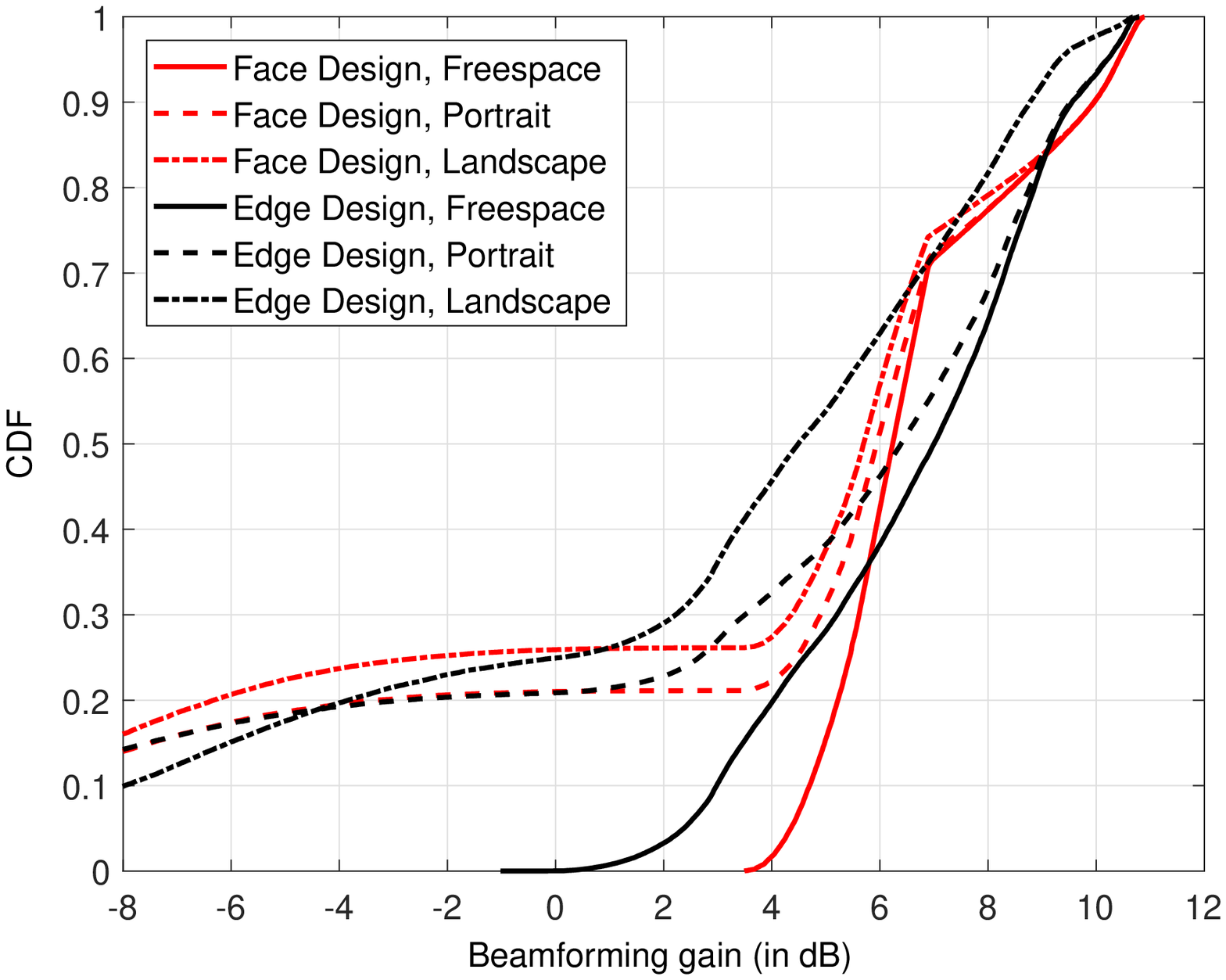}
&
\includegraphics[height=2.3in,width=3.0in]{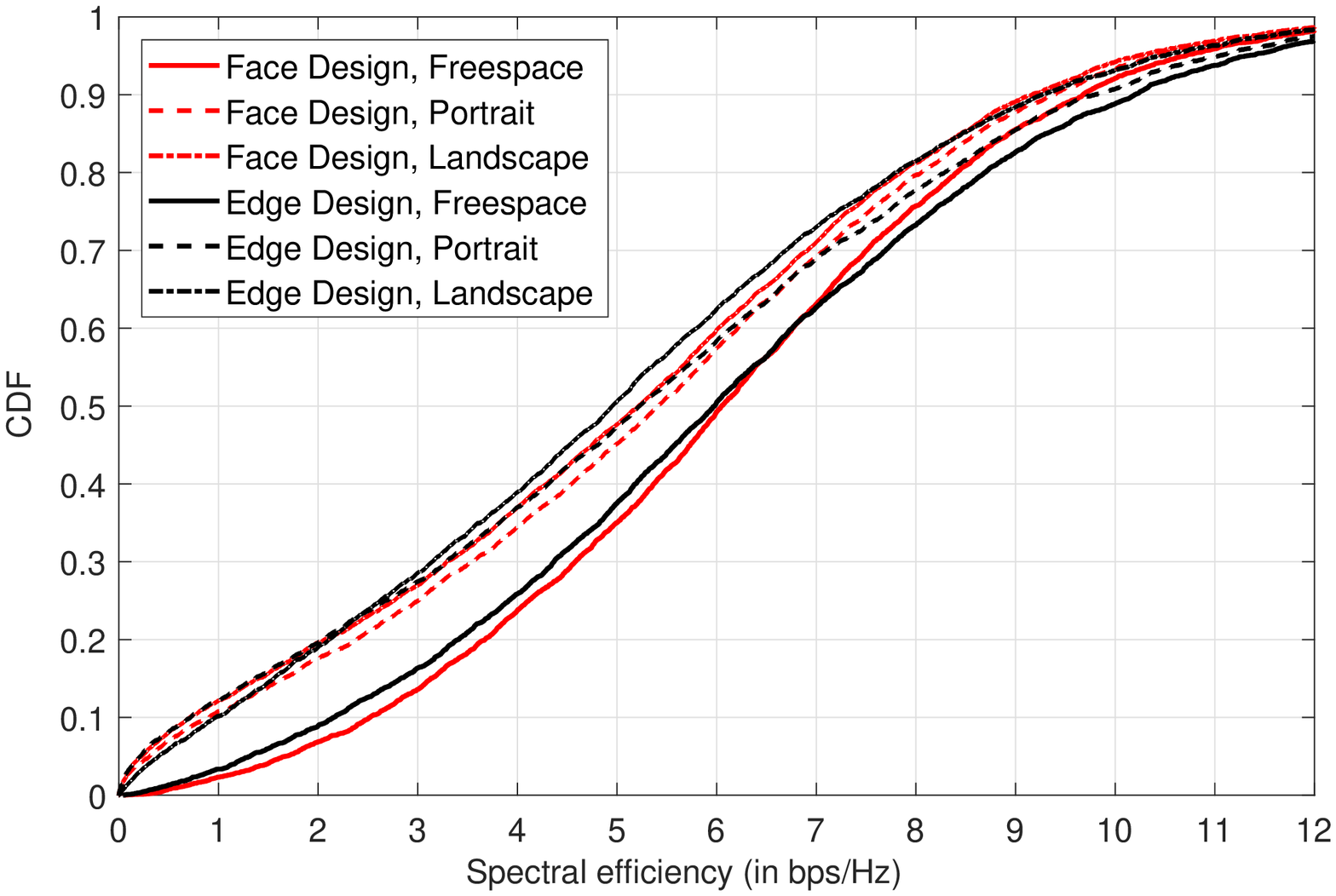}
\\
(a) & (b)
\end{tabular}
\caption{\label{fig_comparison}
Comparative performance between the face and edge designs in Freespace and with blockage (Model 2):
(a) Array gain and (b) Spectral efficiency.}
\end{center}
\end{figure*}

Since the face and edge designs are directly and fairly comparable with each other
(due to the same codebook sizes), Fig.~\ref{fig_comparison}
presents a head-to-head comparison of these designs in Freespace and in Portrait/Landscape
modes with blockage. Blockage Model 2 and the RF/analog beam codebook scheme are used in
all the studies in Fig.~\ref{fig_comparison}. From Fig.~\ref{fig_comparison}(a), we
observe that while both the designs are comparable in the top $20$ percentile points
of the sphere in Portrait mode, the edge design appears to be better (by up to $1.5$ dB)
over the next $35$ percentile points. The face design appears to be better over the
remaining $\approx 20$ percentile points before blockage effects kick in.

While both the face and edge designs are blocked over approximately $21\%$ of the
sphere in the Portrait mode (see~(\ref{eq_CDF_loss_Portrait})), their crossovers can
be explained by the following observations:
Approximate beamforming array gain with the RF codebook at the $70^{\sf th}$ percentile
point for the face and edge designs are $7$ and $8.5$ dB, respectively. Similar
numbers for the $50^{\sf th}$ and $30^{\sf th}$ percentile points are $6.5$ vs.\ $7$ dB and
$5.5$ vs.\ $5$ dB, respectively. This tradeoff arises due to the structure of
antenna arrays ($2 \times 2$ planar arrays and $2 \times 1$ linear arrays in the
face design vs.\ $4 \times 1$ linear arrays in the edge design). The better relative
performance of the edge design over the face design in the middle $35$ percentile points
and its reversal in the next $20$ percentile points is directly a result of the
array gain tradeoffs.

On the other hand, the mismatch between the area blocked with the hand in the
right-hand Landscape
mode (top short edge that is totally blocked in the edge design vs.\ the top front
module that is only partially blocked in the face design) means that the
face design appears to be uniformly better than the edge design (by up to $1.5$ dB).
From these observations, there does not appear to be an overwhelming advantage
(defined as greater than $2$-$3$ dB) for either design suggesting that both designs are
comparable in terms of performance and the choice between them should be decided based
on implementation tradeoffs as described in Sec.~\ref{subsec2_b}.


\begin{figure*}[htb!]
\begin{center}
\begin{tabular}{cc}
\includegraphics[height=2.4in,width=3.2in] {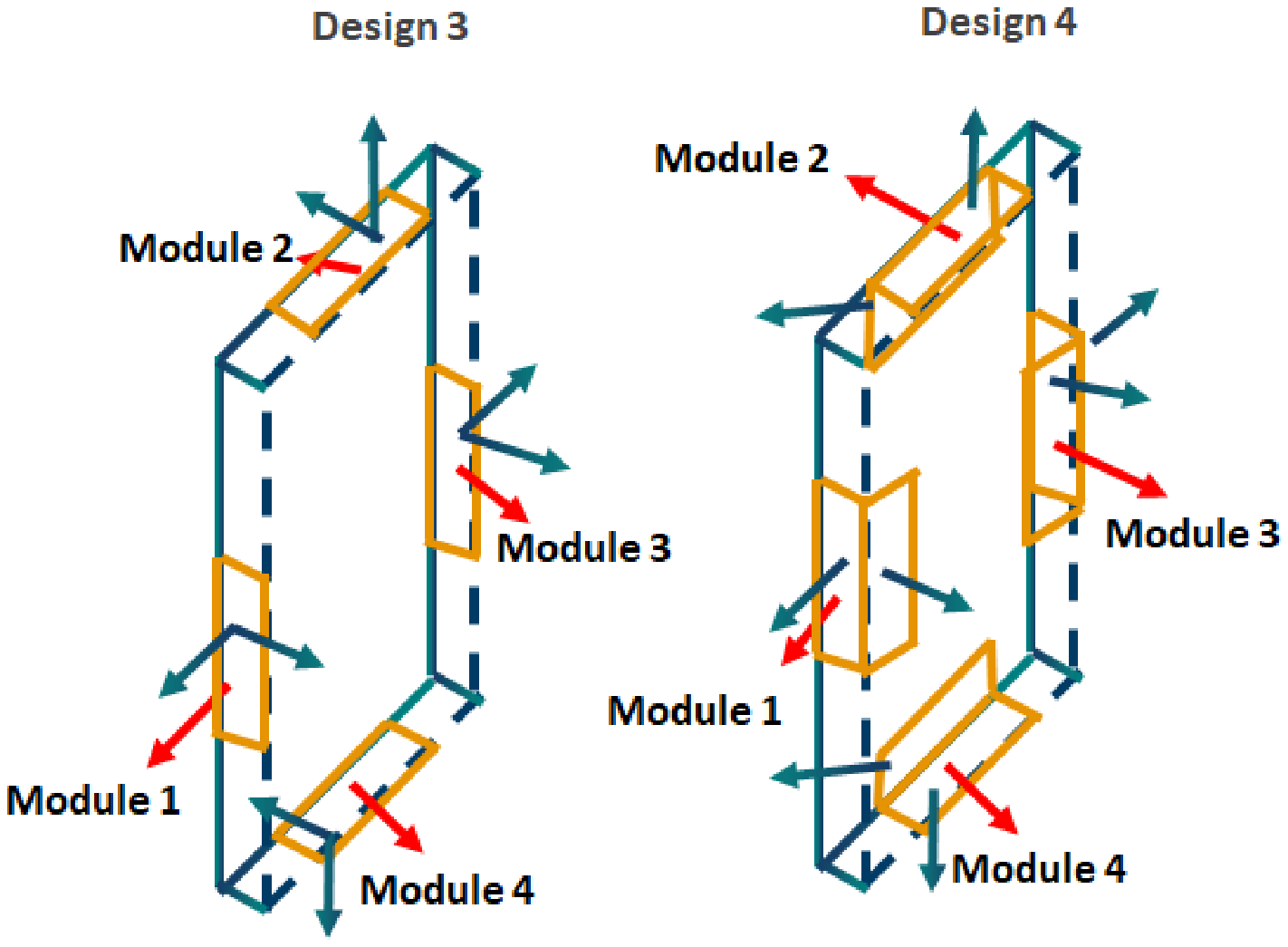}
&
\includegraphics[height=2.3in,width=3.0in]{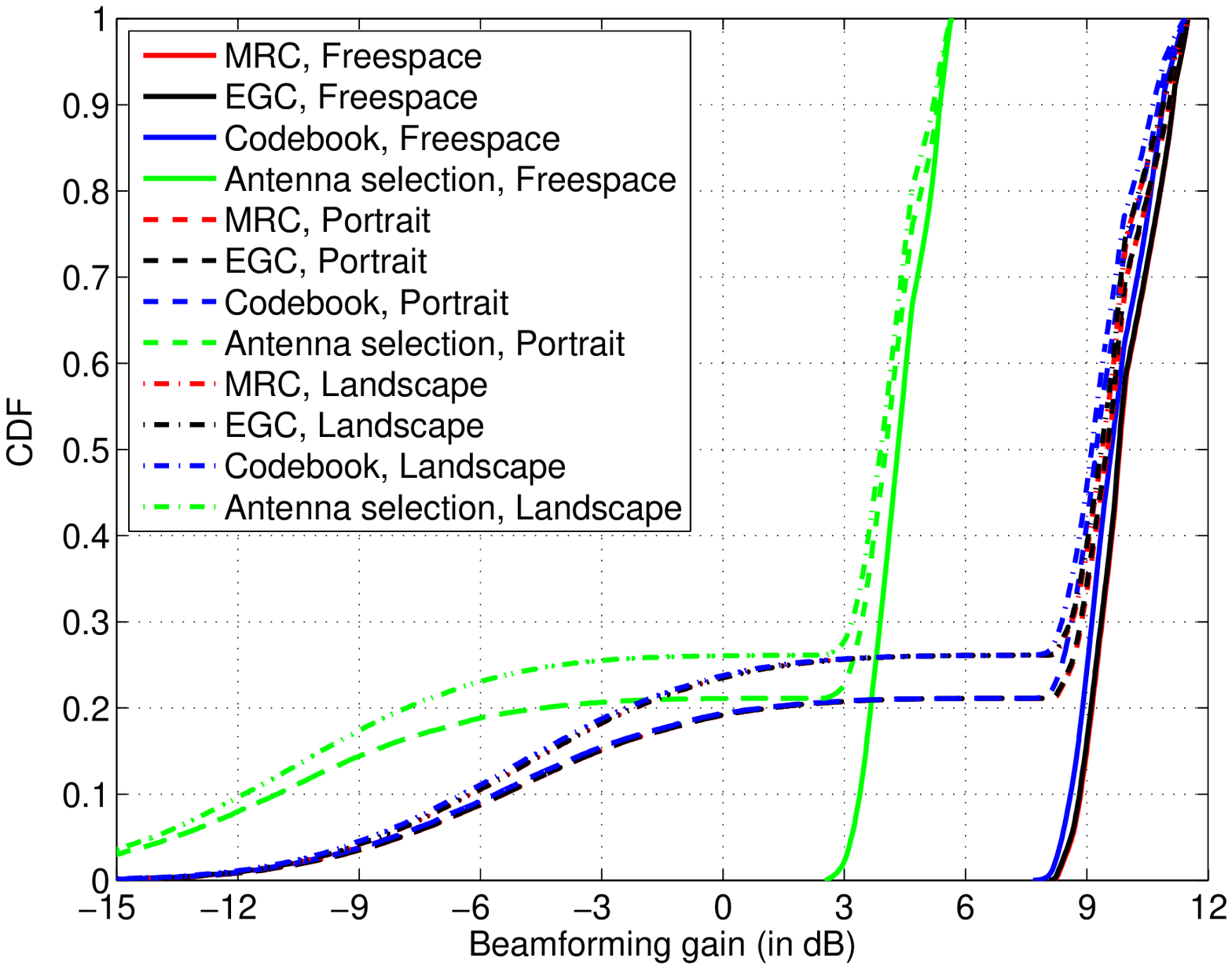}
\\ (a) & (b) \\
\includegraphics[height=2.3in,width=3.0in]{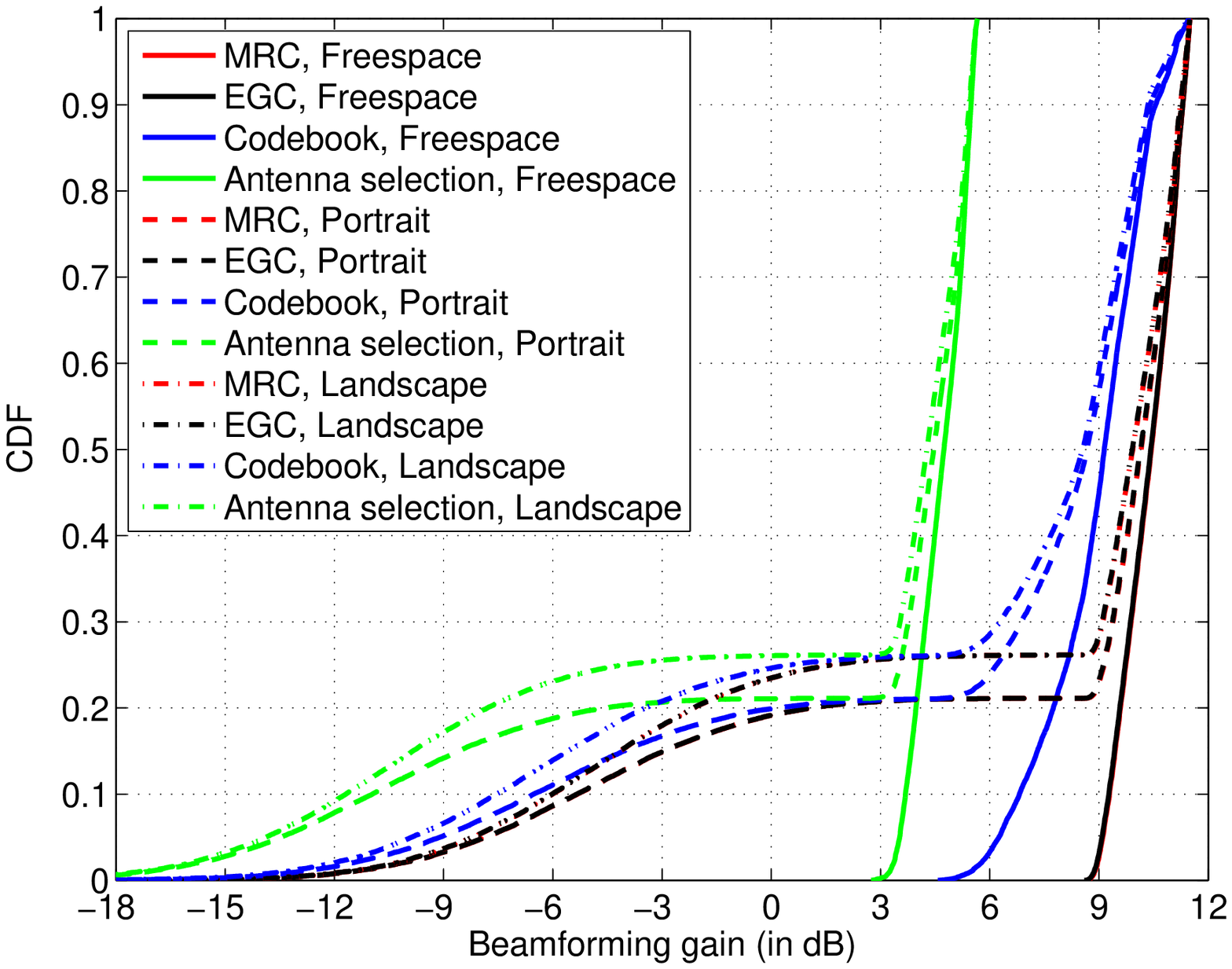}
&
\includegraphics[height=2.2in,width=3.0in]{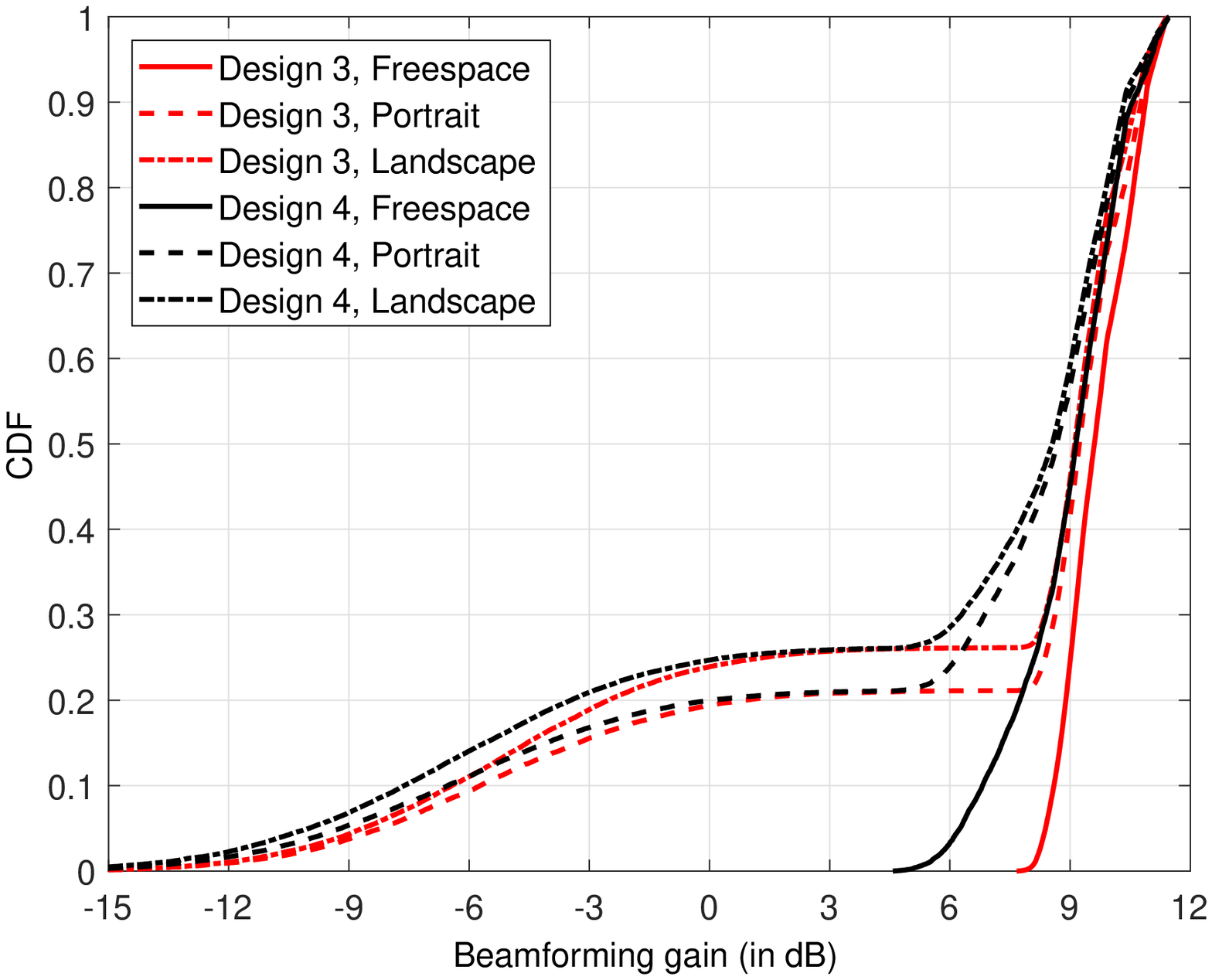}
\\
(c) & (d)
\end{tabular}
\caption{\label{fig_Designs3and4_Model2}
(a) Pictorial illustration of Designs 3 and 4.
Array gain performance with Freespace and blockage in Portrait and Landscape (Model 2) for
(b) Design 3 and (c) Design 4.
(d) Comparative performance between Designs 3 and 4 with blockage Model 2.}
\end{center}
\end{figure*}

Finally, system level simulation studies for an indoor channel environment
with macroscopic fading parameters as described in~\cite{vasanth_tap2018} are performed.
These parameters describe the third and fourth floors of the Qualcomm building in
Bridgewater, NJ.
Table~\ref{table_SLS} illustrates the various parameters used in this comparative study.
In particular, a base-station of size $16 \times 4$ using a size $16$ RF/analog beamforming
codebook (of DFT beams) covering a $120^{\sf o} \times 30^{\sf o}$ coverage area and UEs
according to the face and edge designs with RF/analog beam codebooks as described in
Sec.~\ref{subsec3_b} are used in these studies. Note that these codebooks can be
considered to represent the performance at the end of the P-1/2/3 procedures. The
spectral efficiencies (per layer) in bps/Hz with dual layer
polarization-MIMO transmission/reception at $28$ GHz are illustrated in Fig.~\ref{fig_comparison}(b).
From this study, we observe that the edge and face designs tradeoff performance with each other
in Freespace with the edge design being better in the (approximately) top $50$ percentile
points (by up to $0.3$ bps/Hz or around $1$ dB) and the face design being better in the (approximately)
bottom $50$ percentile points. The trends with blockage are as before, reinforcing the
conclusions made previously.

\begin{table}[htb!]
\caption{System Level Simulation Parameters}
\label{table_SLS}
\begin{center}
\begin{tabular}{|c||c|}
\hline
Metric & Value used \\
\hline
\hline
Base-station antenna dimensions & $16 \times 4$ \\ \hline
Base-station coverage area & $120^{\sf o} \times 30^{\sf o}$ \\ \hline
Base-station codebook size & $16$ \\ \hline
UE coverage area & $360^{\sf o} \times 180^{\sf o}$ \\ \hline
UE codebook size & $24$ \\ \hline
EIRP & $45$ dBm \\ \hline
Bandwidth & $100$ MHz \\ \hline
UE noise figure & $10$ dB \\ \hline
Channel environment & Indoor, \\
& PLE = $3.46$, $\sigma_{\sf SF} = 8.31$ dB \\ \hline
Distance from base-station to UE & $30$ m \\ \hline
No. of clusters & $4$ \\
\hline
\hline
\end{tabular}
\end{center}
\end{table}

\subsection{Generalizations to Other UE Designs}
To study the utility of the above conclusions to other UE designs, we consider two
other popular designs in the literature, as illustrated in Fig.~\ref{fig_Designs3and4_Model2}(a).
These designs are
\begin{itemize}
\item Design 3: A {\em maximalist edge} design with four antenna modules (on four
sides of the UE) with each module made of $4 \times 1$ dual-polarized patch subarrays
and a $4 \times 1$ dipole subarray. While full spherical coverage can be obtained with
patch elements alone, the use of dipole elements provides complementary coverage and
hence, better robustness at the expense of cost associated with more antenna elements
as well as the control circuitry for these elements.

\item Design 4: An {\em L-shaped edge} design with four antenna modules (on four
sides of the UE) with each module being L-shaped and spanning two adjacent sides of
coverage. Each side of coverage is made of $4 \times 1$ dual-polarized patch subarrays
alone.
\end{itemize}
Note that a version of Design 3 has been proposed in~\cite{sony_ericsson_UEdesign2,sony_UEdesign4}
with dipole antennas instead of dual-polarized patches and dipoles, which is a minor design
enhancement. A number of features of Design 4 can be seen in other designs such
as~\cite{ericsson_sony_refUEdesigns,qcom_UEdesign,apple_intel_UEdesign,sony_UEdesign2,sony_ericsson_UEdesign,sony_ericsson_UEdesign2}, as well as~\cite{helander}.

\newcounter{mytempeqncnt1}
\begin{figure*}[htb!]
\normalsize
\setcounter{mytempeqncnt1}{\value{equation}}
\setcounter{equation}{16}
\begin{eqnarray}
J & = & \left[
\begin{array}{ccc}
\frac{ \partial x}{\partial r} & \frac{ \partial x}{\partial \theta} & \frac{ \partial x}{\partial \phi}
\\
\frac{ \partial y}{\partial r} & \frac{ \partial y}{\partial \theta} &
\frac{ \partial y}{\partial \phi}
\\
\frac{ \partial z}{\partial r} & \frac{ \partial z}{\partial \theta} &
\frac{ \partial z}{\partial \phi}
\end{array}
\right] =
\left[ \begin{array}{ccc}
\sin(\theta)\cos(\phi) & r \cos(\theta)\cos(\phi) & -r \sin(\theta)\sin(\phi) \\
\sin(\theta)\sin(\phi) & r \cos(\theta)\sin(\phi) & r \sin(\theta)\cos(\phi) \\
\cos(\theta) & -r\sin(\theta) & 0
\end{array}
\right]  
\label{eq_top1} \\
F(\alpha) & = &
\frac{
\int_{r =0}^R \int_{\theta = 0}^{\pi} \int_{\phi = 0}^{2 \pi}
\indic \left( G_{ {\sf scheme} , {\hspace{0.02in}} {\sf total} } (\theta, \phi)
\leq \alpha \right) r^2 |\sin(\theta)| dr d\theta d\phi  }
{  \int_{r =0}^R \int_{\theta = 0}^{\pi} \int_{\phi = 0}^{2 \pi}
r^2 |\sin(\theta)| dr d\theta d\phi  }
\label{eq_top2}
\\
& = & 
\frac{
\int_{\theta = 0}^{\pi} \int_{\phi = 0}^{2 \pi}
\indic \left( G_{ {\sf scheme} , {\hspace{0.02in}} {\sf total} } (\theta, \phi)
\leq \alpha \right) \sin(\theta) d\theta d\phi } {4 \pi}. 
\label{eq_spherical_CDF}
\end{eqnarray}
\vspace*{4pt}
\hrulefill
\end{figure*}
\setcounter{equation}{\value{mytempeqncnt1}}

Since both these designs have four antenna modules and have far more subarrays ($12$
and $16$, respectively) than either the face or edge designs ($8$ and $6$, respectively),
we use an RF codebook of larger size (size of $48$) than those used with the face and
edge designs (size of $24$). Note that a smaller codebook size with Designs 3 and 4
can lead to coverage holes with poor spherical coverage tradeoffs. For Design 3, we use
$4$ beams for each polarization of the patch and dipole subarray corresponding to $12$
beams per antenna module for a codebook size of $48$. For Design 4, we use $3$ beams
for each polarization of the patch corresponding to $12$ beams per antenna module, also
for a codebook size of $48$. For these two designs,
Figs.~\ref{fig_Designs3and4_Model2}(b)-(c)
present the beamforming array gain comparison with the four beamforming schemes in
Freespace and with hand blockage in Portrait and Landscape modes using Model 2. Similar
to the face and edge designs, we observe that for both Designs 3 and 4, the RF/analog
beam codebooks are within $1$-$2$ dB of the MRC/EGC performance suggesting the goodness
of the codebook design principles. In particular, the worst-case points of the codebook's
performance in Freespace 
are $3$ dB and $6$ dB away (which is better than the face and edge designs) from the peak
gain for these designs. As intuitively expected
from a co-phasing with four antennas in either design, single antenna selection is
approximately $5$-$6$ dB worse than MRC/EGC. Blockage tradeoffs for both the designs are
similar to those described earlier for the face and edge designs.

Since Designs 3 and 4 are directly and fairly comparable with each other,
Fig.~\ref{fig_Designs3and4_Model2}(d) provides a comparison across these two designs.
From this study, we observe that Design 3 has a universally (albeit slightly) better
performance (in Freespace as well as with blockage) over Design 4. This plot suggests
that the use of dipoles over patches that scan the other side of the L can result in a
better performance for diversity. Thus, the use of the appropriate/correct antenna
modules is crucial for good performance in millimeter wave systems.

\section{Concluding Remarks}
\label{sec5}
The focus of this paper has been on the study of spherical coverage CDF of two
popular millimeter wave UE designs with real impairments such as hand blockage.
The designs considered in this paper correspond to a face design and
an edge design, respectively. For our studies, we considered four types of
beamforming schemes (MRC, EGC, RF/analog beam
codebook and antenna selection) with two types of blockage models (3GPP blockage
model and a modified version of the 3GPP model). From our studies, we established
the overhead of beam training as being the key determinant (and not the
``theoretical'' capabilities enabled with multiple antenna modules) for robust
spherical coverage performance. {\em That is, it is not merely sufficient that the UE
is packed with a large number of antenna modules, but that the subarrays in these
modules have to be scanned/learned with an appropriately designed beam codebook in a
practical implementation.} Further, the size of a good codebook has to scale with
the number of antenna modules and can render the coverage gains unrealizable from
a practical standpoint. From this view, we established the goodness of the edge UE
design that also has other additional advantages such as low cost and power consumption,
implementation ease, and minimal exposure related challenges~\cite{qualcomm_mmw_modules}.
Table~\ref{table_designtradeoffs} provides a broad overview and summary of
the major issues with these designs.

\begin{table*}[htb!]
\caption{Broad Tradeoffs for the Different UE Designs}
\label{table_designtradeoffs}
\begin{center}
\begin{tabular}{|c||c|c|}
\hline
Issue of interest & Face design & Edge design  \\
\hline
\hline
\multicolumn{3}{|c|} {Design/cost/regulatory tradeoffs} \\
\hline
Mounting problems & High & Low \\ \hline
No.\ of antenna modules & $2$ & $3$ \\ \hline
Exposure-related challenges & Major & Minor \\ \hline
Dipole-related pros/cons & Yes & No \\
\hline
\hline

\multicolumn{3}{|c|} {Beam scanning complexity} \\
\hline
Worst-case initial acquisition overhead & $80$ ms & $60$ ms \\
\hline
Beam localization with $2$D arrays & Possible with patches & Not possible \\
\hline
\hline

\multicolumn{3}{|c|} {Beamforming performance} \\
\hline
Link budget needed to overcome penetration & More & Less \\ \hline
Freespace performance at cell center & Poor & Better \\ \hline
Freespace performance at cell edge & Better & Poor \\ \hline
Performance with Portrait blockage & Better from $55$th to $75$th percentiles &
Better from $20$th to $55$th percentiles \\ \hline
Performance with Landscape blockage & Universally better &
Universally poorer \\
\hline
\hline
\end{tabular}
\end{center}
\end{table*}

\ignore{
\begin{table*}[htb!]
\caption{Broad Tradeoffs for the Different UE Designs}
\label{table_designtradeoffs}
\begin{center}
\begin{tabular}{|c||c|c|}
\hline
Issue & Face Design & Edge Design  \\
\hline
\hline
No.\ of antenna modules & $2$ & $3$ \\
\hline
Beam scanning & 2-D for patches & 1-D \\
& 1-D for dipoles & \\
\hline
Mounting challenges & Yes & No \\
\hline
Penetration challenges & Yes & No \\
\hline
Exposure-related challenges & Major & Minor \\
\hline
Dipole pros/cons & Yes & No \\
\hline
\hline
\end{tabular}
\end{center}
\end{table*}

}

That said, this work has barely scratched the surface in terms of coupling different
practical/commercial UE design challenges with their system level impacts. In fact,
this work has exposed the challenges of good UE designs for millimeter wave transmissions,
{\em which are quite unlike those of sub-$6$ GHz systems.} More work
is necessary to understand the impact of optimal codebook construction/beamforming
schemes on spherical coverage, incorporating priors (e.g., base-station downtilt,
UE modalities, etc.) on spatial angles in spherical coverage studies, initial acquisition
vs.\ steady-state performance tradeoffs, etc. Further work is also necessary in incorporating
metrics that capture multiple layer/RF chain performance and establishing the structure
of optimal UE designs for such metrics. Leveraging antenna response functions in a UE
design incorporated with practical display and frame materials such as glass, plastic, ceramic,
etc., are also important in future studies.

\appendix

\subsection{Computing Spherical Coverage CDF}
\label{app_spherical}
Let $G_{ {\sf scheme} , {\hspace{0.02in}} {\sf total} } (x,y,z)$ denote the total array
gain (over both polarizations) of a certain beamforming scheme at a point $(x,y,z)$
represented in the ${\sf X}$-${\sf Y}$-${\sf Z}$ Cartesian coordinate system. Then, the
CDF of spherical coverage evaluated at $\alpha$ over a sphere
of radius $R$ is given as
\begin{eqnarray}
F(\alpha) = \frac{ \iiint \indic \left( G_{ {\sf scheme} , {\hspace{0.02in}} {\sf total} } (x,y,z)
\leq \alpha \right) dx dy dz } { \iiint dx dy dz }
\label{eq_Falpha}
\end{eqnarray}
where $\indic(\bullet)$ denotes the indicator function of the underlying variable. The
differential element in the Cartesian coordinate system is transformed to the differential
element in the spherical coordinate system (with $x = r\sin(\theta) \cos(\phi)$,
$y = r\sin(\theta) \sin(\phi)$ and $z = r\cos(\theta)$) as
\begin{eqnarray}
dx dy dz = {\cal J} dr d\theta d\phi
\end{eqnarray}
where ${\cal J} = |{\sf det}(J)|$ with $J$ denoting the Jacobian matrix of the
transformation, as described in~(\ref{eq_top1}) at the top of the previous page,
resulting in ${\cal J} = r^2 |\sin(\theta)|$. With this,~(\ref{eq_Falpha}) transforms
to the description in~(\ref{eq_top2})-(\ref{eq_spherical_CDF}) at the top of the
previous page. It is critical to
note the scaling factor $\sin(\theta)$ in~(\ref{eq_spherical_CDF}) which
reduces the weightage of points at the poles (where $\theta = 0$ and $\pi$)
and increases the weightage of points at the equator (where $\theta = \pi/2$).
\qed

{\vspace{-0.05in}}
\bibliographystyle{IEEEbib}
\bibliography{newrefsx2}

\end{document}